\documentclass[twoside, a4paper]{article}
\usepackage[latin1]{inputenc}
\usepackage{amssymb}
\usepackage{epsfig}

\title{State Space Reconstruction Parameters in the Analysis of Chaotic Time Series - the Role of the Time Window Length}
\author{D.~Kugiumtzis \\
        {\em Department of Informatics, University of Oslo,} \\
        {\em P.O.Box 1080 Blindern, N-0316 Oslo, Norway}}
\date{}

\begin{document}
\maketitle

\newcommand{\be}{\begin{equation}}
\newcommand{\ee}{\end{equation}}
\newcommand{\upint}[1]{\mbox{$\lceil #1 \rceil$}}

Key word index: {\em time series, chaos, state space reconstruction, 
                     correlation dimension}

\begin{abstract}
The most common state space reconstruction method in the analysis
of chaotic time series is the Method of Delays (MOD). Many techniques have 
been suggested to estimate the parameters of MOD, i.e. the
time delay $\tau$ and the embedding dimension $m$. We discuss the 
applicability of these techniques with a critical view as to their
validity, and point out the necessity of determining the overall 
time window length, $\tau_w$, for successful embedding.
Emphasis is put on the relation between $\tau_w$ and the dynamics 
of the underlying chaotic system, and we suggest to set 
$\tau_w \geq \tau_p$, the mean orbital period; $\tau_p$ is approximated 
from the oscillations of the time series. The procedure is assessed using the 
correlation dimension for both synthetic and real data. For clean synthetic 
data, values of $\tau_w$ larger than $\tau_p$ always give good results 
given enough data and 
thus $\tau_p$ can be considered as a lower limit ($\tau_w \geq \tau_p$). For 
noisy synthetic data and real data, an upper limit is reached for
$\tau_w$ which approaches $\tau_p$ for increasing noise amplitude. 

\end{abstract}

\section{Introduction}

State space reconstruction is the first step in non-linear time series 
analysis of data from chaotic systems including estimation of invariants 
and prediction. For a recent review of these topics see \cite{Kugiumtzis94} 
and \cite{Lille94}. Reconstruction consists of viewing a time series 
$x_k = x(k\tau_s)$, $k = 1, \ldots ,N$  
in a Euclidean space ${\mathbb{R}}^m$, where $m$ is the {\em embedding dimension}
and $\tau_s$ is the sampling time. 
Doing this, we hope that the points in ${\mathbb{R}}^m$ form an 
attractor that preserves the topological properties of the original
unknown attractor. A standard way to 
reconstruct the state space is the Method of Delays (MOD).
Using MOD,  each $m$-dimensional embedding vector is formed as
$\mathbf x_k= [x_k, x_{k+\rho}, \ldots, x_{k+(m-1)\rho}]^T$
where $\rho$ is a multiple integer of $\tau_s$ so that the 
{\em delay time} $\tau$ equals $\rho \tau_s$ 
\cite{Packard80}. The $m$ coordinates of each point $\mathbf x_k$ are 
samples from the time series (separated by a fixed $\tau$) covering a 
{\em time window} of length $\tau_w = (m-1) \tau$ (or $\tau_w = (m-1) \rho$ 
as multiple of $\tau_s$). 

The fundamental theorem of reconstruction, introduced first by Takens
\cite{Takens81} \footnote{Similar work was made independently in 
\protect\cite{Mane81}.} and extended more
recently in \cite{Sauer91}, gives no restriction on $\tau$ while for $m$
states the sufficient (but not necessary) condition 
$m \geq 2 d + 1$, where $d$ is the fractal dimension of the underlying
attractor \footnote{Actually, Takens' condition uses $\upint d$ 
instead of $d$, the topological dimension, i.e. the lower integer 
greater than $d$. The use of $d$ in the inequality has been
established in \cite{Sauer91} allowing lower values for m.}.
Takens' theorem is valid for the case of infinitely many noise-free
data. In practice, however, with a limited number of possibly noisy 
observations, the selection of $\tau$ and $m$ is rather important for 
the quality of the reconstruction. Many methods have been suggested 
for estimating these parameters, but they are all empirical in nature
and do not -- as we show -- necessarily provide appropriate estimates.
This is a rather typical situation regarding state space reconstruction
in general. 

While there will always be uncertainties related to 
reconstruction from real data, it is still important to try to 
improve 
the procedures. We suggest $\tau_w$ as an 
independent parameter instead of focusing on the interrelated parameters
$\tau$ and $m$ of MOD. 
The time window length is of particular importance since it determines, in 
a certain sense, the amount of information passed from the time series
to the embedding vectors. For a given $\tau_w$, one may then select a
sufficiently large $m$.
Suggestions for the selection of $\tau_w$ have been 
made in \cite{Broomhead86}, \cite{Caputo86}, \cite{Albano88}, \cite{Grassberger91},
\cite{Gibson92} and \cite{Rosenstein93} but to our knowledge there has 
been little systematic work regarding this parameter. We give procedures 
for estimating $\tau_w$ from the signal. Only time series 
from continuous systems are treated. For discrete systems, one typically 
sets $\rho=1$, reducing the number of parameters to one -- 
the embedding dimension, since $\tau_w = m-1$. 

The quality of the reconstructions is assessed using the correlation 
dimension \cite{Grassberger83a}. The resulting 
reconstructions may not be the most suitable for other purposes such
as estimation of Lyapunov exponents and prediction. However,
with improved reconstructions for dimension estimation it is likely that 
the technique will be valuable also in other cases.
 
In section \ref{MODparameters}, we discuss several of the  
methods suggested up to now for estimating $\tau$ and $m$ in MOD 
and comment on the underlying ideas as well as on the 
validity of the results. In section \ref{tau_w}, we establish the role of
$\tau_w$ in reconstruction and give simple ways to estimate it. 
Finally, in section \ref{cordimtau_w}, the correlation dimension 
is used to assess the proposed procedure using noise-free and
noise-corrupted synthetic data as well as real data.

\section{Suggested methods for estimating the MOD-parameters}
\label{MODparameters}

A very helpful approach in visualizing the reconstruction problem is to
consider the reconstruction as an orthogonal projection from some high
$p$-dimensional state space onto an $m$-dimensional subspace defined by 
the $m$ coordinates of the reconstructed vectors. Defining the linear
mapping $B:{\mathbb{R}}^p \longrightarrow {\mathbb{R}}^m$, from each 
$p$-dimensional vector ${\mathbf x}_k^p$ to an $m$-dimensional 
vector ${\mathbf x}_k^m$, we have ${\mathbf x}_k^m = B {\mathbf x}_k^p$,
where the rows of the $m \! \times \! p$ matrix $B$ are orthonormal. 
The $p$ coordinates of ${\mathbf x}_k^p$ are actually all the samples 
in the time window $\tau_w$ 
and in the case of MOD, where $p - 1 = \tau_w = (m-1) \rho$, the 
$m$ coordinates of the projected subspace are every $\rho$'th sample 
starting with the first, i.e. each row of $B$ has one 1 and $p-1$ zeros. 
Obviously, one 
can find other $m$-dimensional subspaces using a smaller $\rho$ 
(which may not cover the whole $\tau_w$). Using $\rho=1$ results 
in an unfavorable reconstruction if 
the time series is densely sampled because then the attractor lies on the 
diagonal in ${\mathbb{R}}^m$. (The successive samples differ very little 
from each other.) In such a projection we utilize only 
the $m$ first samples of $\tau_w$. Other projections may be considered 
such as the one employed in the Singular Spectrum Approach 
(SSA) \cite{Broomhead86}. This method yields first a transformation of 
the natural coordinate system to another orthogonal system, ranking the 
$p$ new directions according to the variance they explain, followed by a 
projection onto the $m$ first directions. The rows of the $B$ matrix 
are then the first $m$ eigenvectors of the $p \! \times \! p$ sample 
covariance matrix of the embedding vectors.  
The reconstruction viewed as a projection from the hyperspace 
determined by $\tau_w$ reveals the importance of this parameter. 
For MOD, the subspace is defined 
completely by the parameters $\tau$ (or $\rho$) and $m$ and for SSA by 
$p$ and $m$. 
 
Certain statements supporting current methods for estimating $\tau$ and
$m$ have been widely accepted and almost adopted as axioms. We do 
not intend to question all the existing methodology on MOD state
space reconstruction, but feel that a discussion is needed regarding 
the guidelines used to choose the parameters. 

\subsection{Comments on the selection of the delay time}

Consider first $\tau$ and the two following widely accepted criteria:   
\begin{enumerate}
  \item The reconstructed attractor must be expanded from the diagonal
        (implying that $\tau$ should not be too small) but not too much
        so that it folds back (implying that $\tau$ should not 
        be too large).
  \item The components of the vector $\mathbf x_k$ must be uncorrelated.
\end{enumerate}  

\begin{figure}[htb] 
\hspace{-5mm}
\centerline{\hbox{\psfig{file=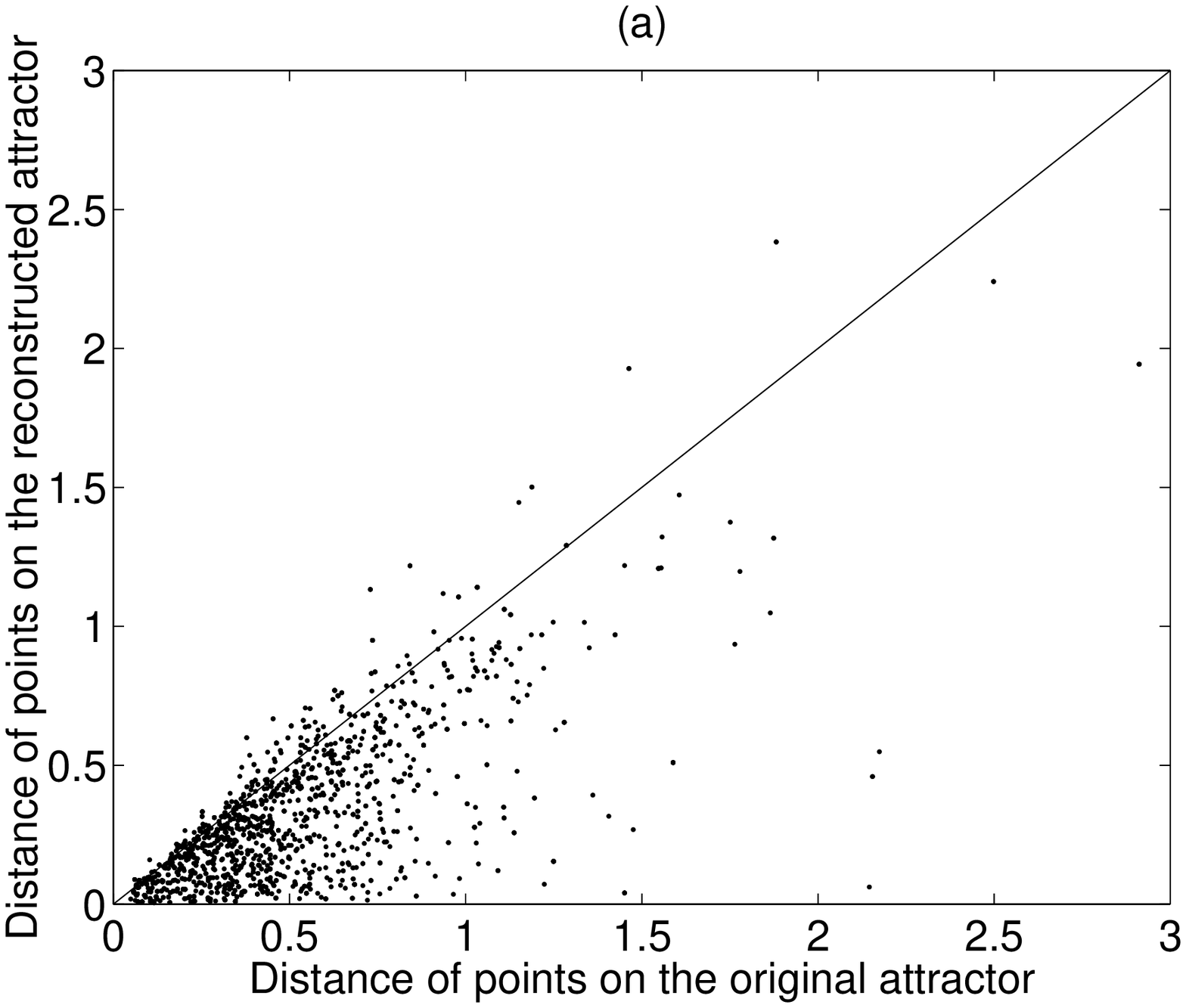,height=5cm,width=5cm}
                 \psfig{file=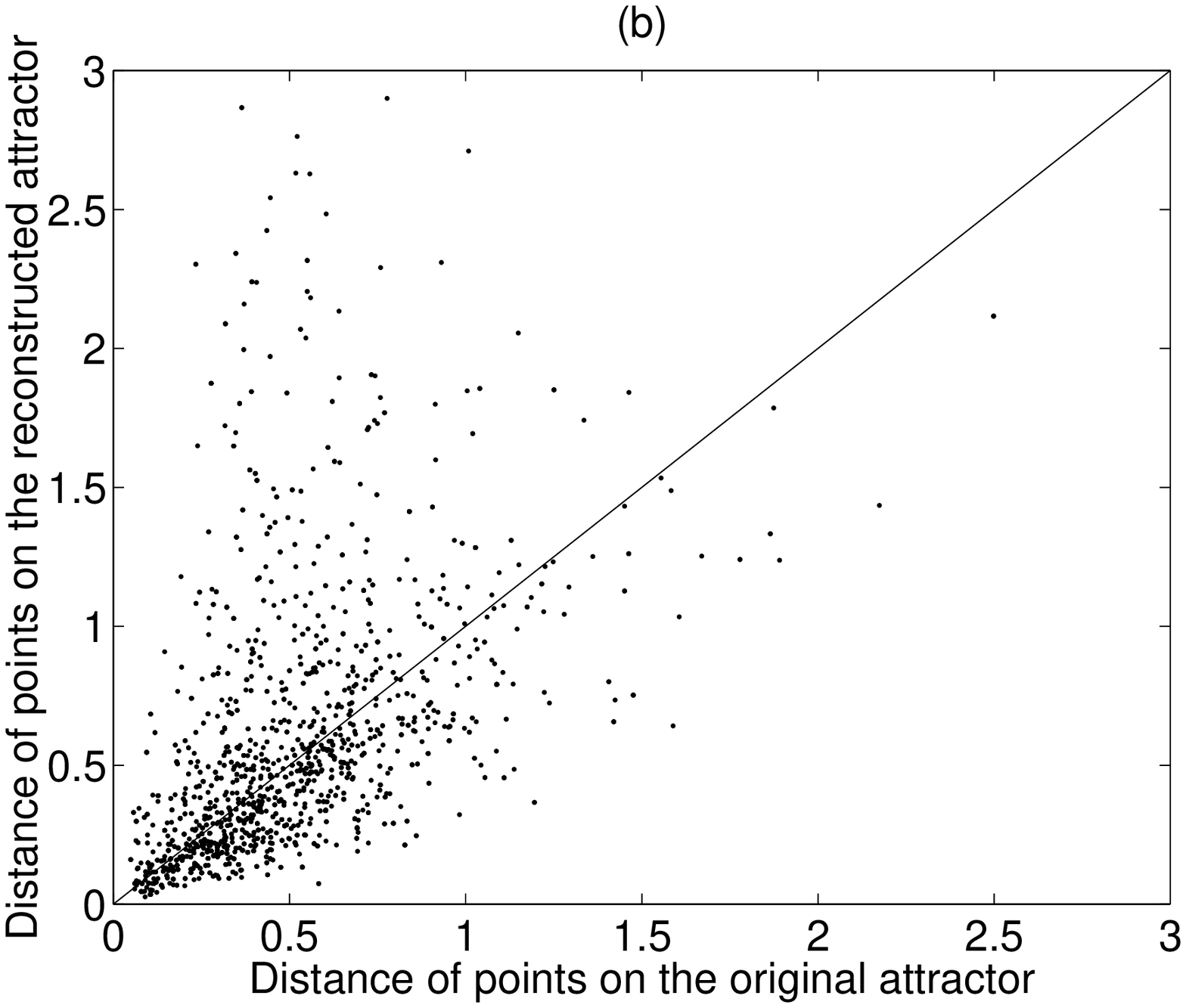,height=5cm,width=5cm}
                 \psfig{file=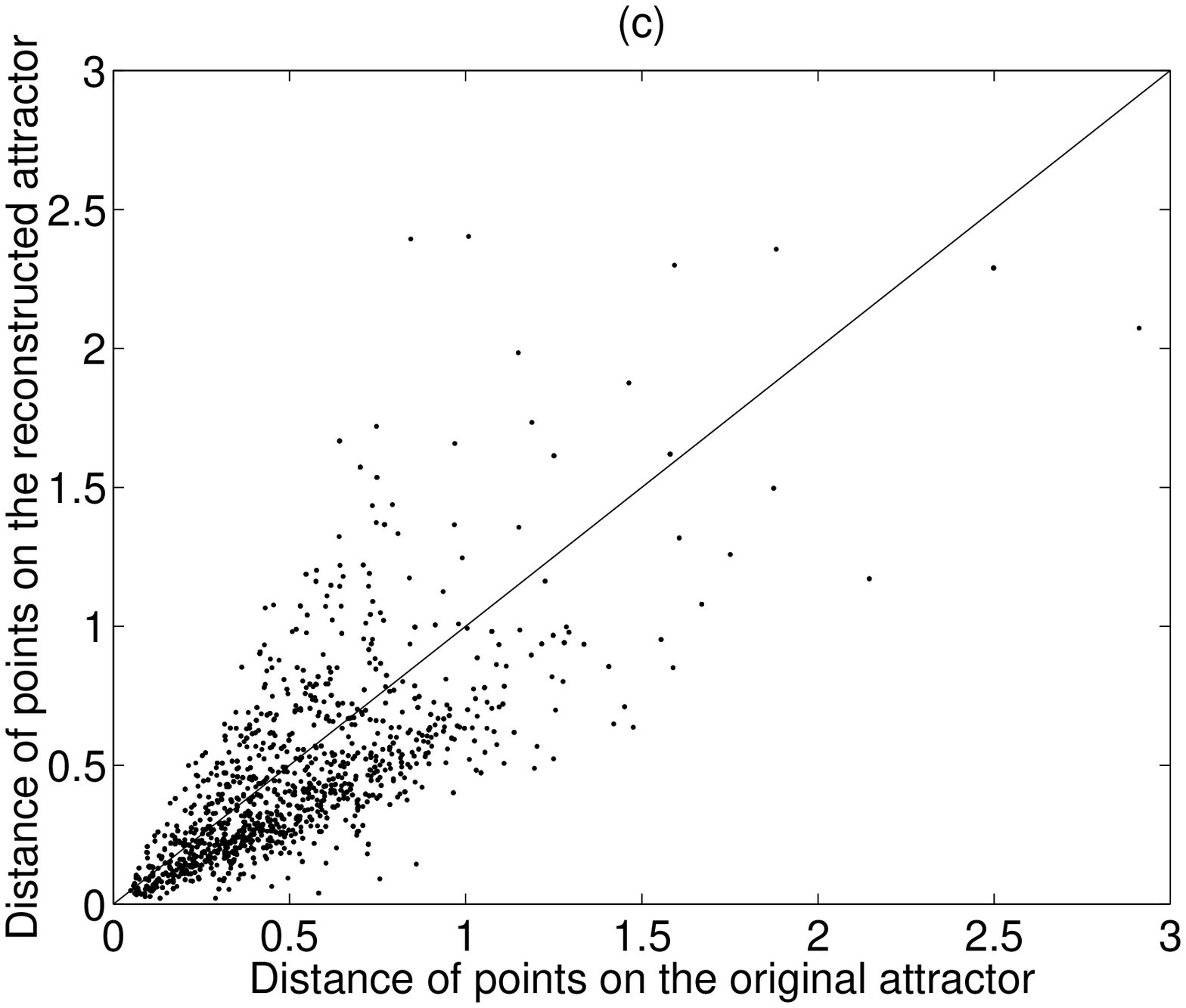,height=5cm,width=5cm}
           }}
\caption{Correlation diagrams of the distances of points on the
         original attractor ($x$-axis) and on the reconstructed 
         attractor ($y$-axis) for the Lorenz system. Results are shown
         for 10\% of the 20000 data points sampled with $\tau_s = 0.01$
         time units. For each point on the original attractor the 
         distance from its nearest neighbor is computed
         and keeping track of the time indices the distance of 
         the corresponding points on the reconstructed attractor
         is then found.
         The attractor is reconstructed with MOD, $m=3$ and $\rho=1$ in 
         (a), $\rho=18$ in (b), and $\rho=9$ in (c).}
\label{fig1}
\end{figure}

Note the similarity of the two criteria: increasing $\tau$ expands the 
attractor from the diagonal and the components get less correlated; beyond
some range of $\tau$, folding may occur and the components again get 
correlated. These goals are intuitively reasonable for $m=2$, while 
the generalization to a larger $m$ is not always straightforward as we show 
below. Many methods based on geometric properties seek the $\tau$ that 
makes the attractor cover the largest region or expands it maximally from 
the diagonal \cite{Buzug92a}, \cite{Rosenstein93}, \cite{Kember93}. 
However, the goal of stretching the attractor from the diagonal to get 
``good'' reconstructions is based rather on empirical than theoretical 
grounds. In theory, a good reconstruction means near topological 
equivalence of the reconstructed attractor to the original one.
One way to assess topological equivalence is to check whether stretching 
and folding are proportionally the same in the two attractors. In practice,
this is done by checking whether the inter-distances of points remain
proportionally the same in the two attractors or, alternatively, 
by checking whether nearby points on the original attractor remain 
relatively close on the reconstructed attractor. 
This last property is not always preserved when we expand the 
attractor from the diagonal, even for proper expansions according
to the two above criteria. We show this for
the Lorenz system \cite{Lorenz63} in Fig.\,\ref{fig1}. 
Fig.\,\ref{fig1}a shows that when $\tau$ is very small 
($\tau=0.01$) the reconstructed attractor 
lies almost on the diagonal and the points are generally getting 
closer than the corresponding points on the original attractor. One 
expects that this problem is resolved when we expand the attractor 
sufficiently ($\tau=0.18$ which gives the minimum of the so-called
mutual information -- see below). But the 
opposite phenomenon is observed instead as shown in Fig.\,\ref{fig1}b, 
i.e. points that are close on the original attractor become more distant 
on the reconstructed attractor. Further, we show in Fig.\,\ref{fig1}c
that the distances are more balanced for the reconstruction with 
a comparably small value of $\tau$
($\tau=0.09$) which is not apparent from the two above criteria.
The point we want to infer from this remark is that there is not 
necessarily a meaningful answer to the question:
Why should we seek the $\tau$ that gives sufficient expansion from the 
diagonal? Expansion {\it per se} does not guarantee a 
configuration of the reconstructed attractor closer to the original one. 

\begin{figure}[htb] 
\centerline{\hbox{\psfig{file=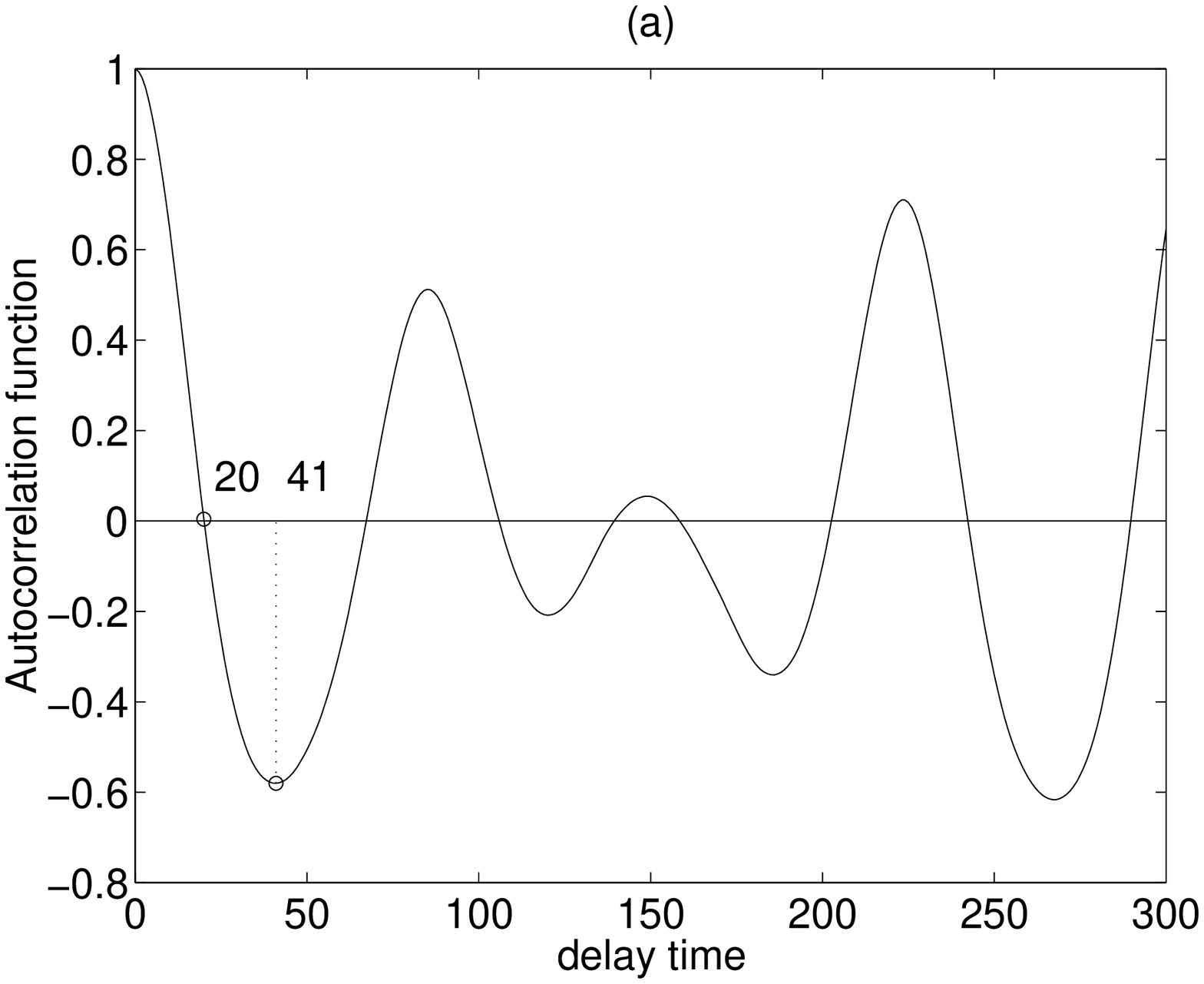,height=5cm,width=5cm}
                 \psfig{file=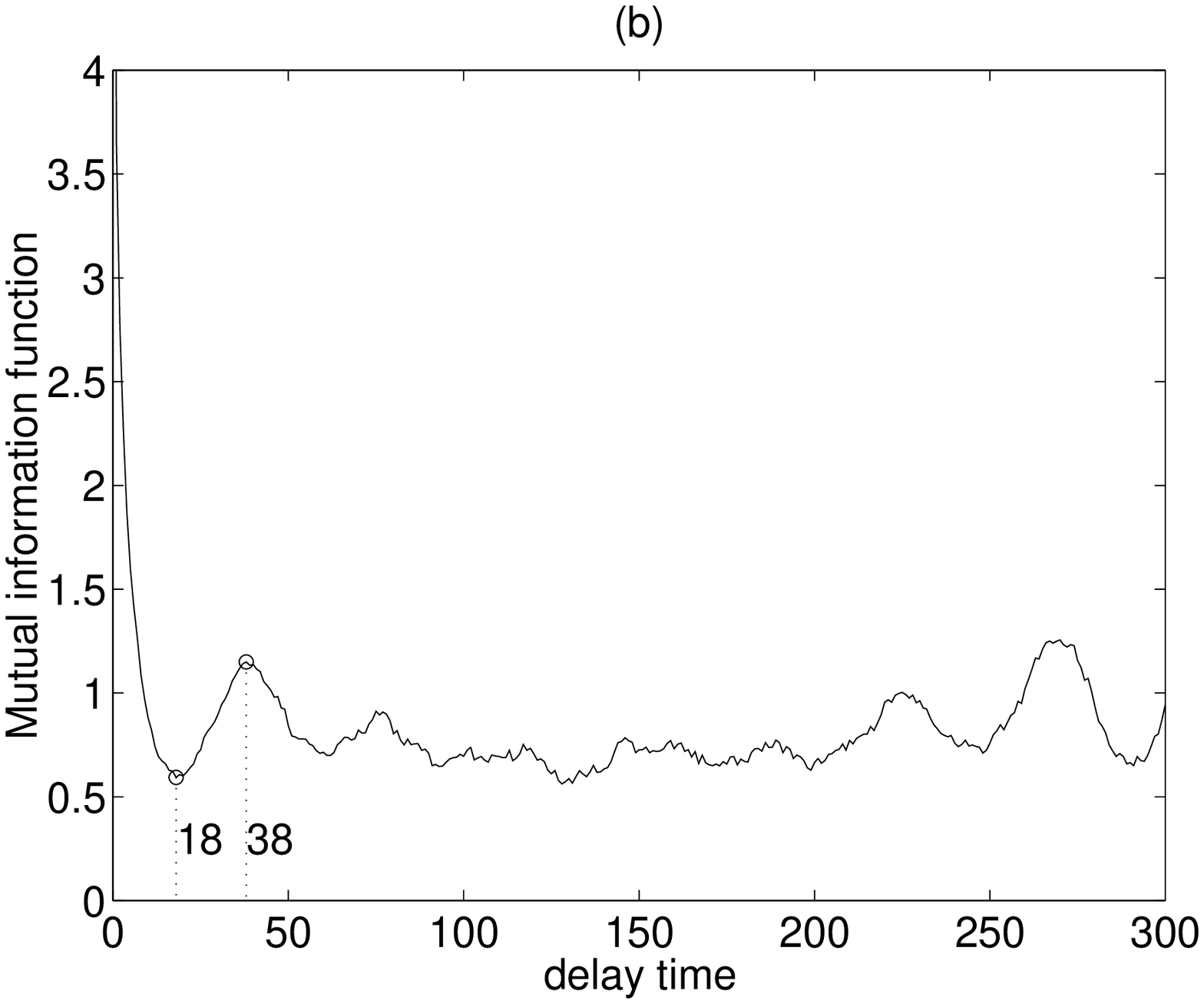,height=5cm,width=5cm}
           }}
\caption{Autocorrelation function $R(\tau)$ in (a) and mutual 
         information $I(\tau)$ in (b)
         for a time series of 10000 data measured from the Taylor-Couette
         experiment in the chaotic regime. 
         Note the approximate matching of the zeros of $R$ to minimums of $I$ and
         the extremes of $R$ to maximums of $I$ indicating a dominant 
         linear correlation. Moreover, note that the first decorrelation time
         is for $\rho \sim 20$ while for $\rho \sim 40$ there is maximum 
         correlation.} 
\label{fig2}
\end{figure}

Concerning the second criterion, the estimates for $\tau$ are based 
either on linear decorrelation, choosing $\tau$ such that
$R(\tau)=0$, where $R$ is the autocorrelation function\footnote{Other 
values of $R(\tau)$ such as $R(\tau)=1/e$ have also been suggested 
but used little in applications, e.g. see \protect\cite{Tsonis92}.},
or general decorrelation choosing $\tau$ to be the first minimum
of the mutual information $I(\tau)$ as developed in \cite{Fraser86}. These
two methods guarantee decorrelation (linear or general) between two 
successive components
$x_k$ and $x_{k+\tau}$ of the reconstructed vector $\mathbf x_k$.
But even if $x_k$ and $x_{k+\tau}$ are uncorrelated and 
$x_{k+\tau}$ and $x_{k+2\tau}$ are uncorrelated, it does not follow
that $x_k$ and $x_{k+2\tau}$ are also uncorrelated. 
As an example, we show in Fig.\,\ref{fig2} $R$ and $I$ for
a time series from the Taylor-Couette experiment in the chaotic 
regime \cite{Brandstater87} which exhibits strong decorrelation for 
some lag $\tau$ and strong correlation for lag $2\tau$. 
We believe that the behavior of the correlation functions 
in Fig.\,\ref{fig2}
are often met in applications since chaotic time series
from low dimensional systems frequently show pseudo-periodicities.

One may be confronted also with other problems attempting to estimate 
$\tau$:
the autocorrelation function may get approximately zero only after an 
extremely long time, as
for the $x$-variable of the Lorenz system, or the mutual information 
may not have a clear minimum, as is the case with the physiological data 
used below. 

\subsection{Comments on the selection of the embedding dimension}

The standard way to find $m$ is to use some criterion which the
geometry of the attractor must meet and check for which embedding dimension $m^*$ 
this is fulfilled as the attractor is embedded in successively 
higher dimensional spaces.
Then $m^*$ is the lowest embedding dimension to be used for reconstruction.
Obviously, in estimating $m$, $\tau$ is fixed when MOD is used.

Among different geometrical criteria (including also the correlation 
dimension), the most popular seems to be the method of ``False Nearest 
Neighbors'' (FNN) developed in \cite{Kennel92} and enhanced recently in 
\cite{Kennel95}. The rationale behind this method has also been discussed
in \cite{Liebert91} and \cite{Aleksic91}. 
This criterion concerns the fundamental condition of no 
self-intersections of the reconstructed attractor. The original attractor 
lies on a smooth manifold of dimension $\upint d$.
Self-intersections of the reconstructed attractor indicate that it does not
lie on a smooth manifold and thus the reconstruction is not successful.
The condition of no self-intersections states that if 
the attractor is to be reconstructed successfully in ${\mathbb{R}}^m$, then all
neighbor points in ${\mathbb{R}}^m$ should also be neighbors in ${\mathbb{R}}^{m+1}$. 
The method checks the neighbors in successively higher embedding 
dimensions until it finds only a negligible number of false neighbors 
when increasing the dimension from $m^*$ to $m^*+1$. This $m^*$ is chosen as
the lowest embedding dimension that gives reconstructions without 
self-intersections.
However, the fact that the distances between neighboring points 
do not change when measured in ${\mathbb{R}}^m$ and in ${\mathbb{R}}^{m+1}$, 
does not necessarily mean that these points are also true neighbors on the 
original attractor. 

\begin{figure}[htb] 
\centerline{\psfig{file=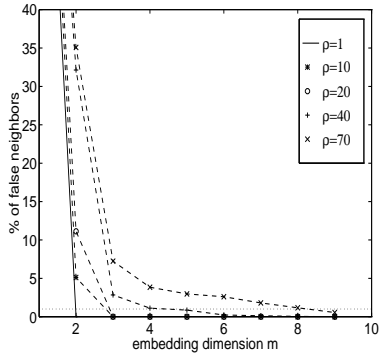,height=5cm,width=5cm}}
\caption{Plot of the percent of false neighbors detected as
  the embedding dimension is increased for different values of $\tau$. 
  The algorithm of FNN has been implemented for a time 
  series of 10000 samples of the $x$ variable of the Lorenz system. The 
  different curves correspond to the time delays given in the legend 
  as multiple of the sampling time $\tau_s = 0.01$. The horizontal 
  stippled line shows the 1\% level of false neighbors which is often
  used as the discriminative threshold value.}  
\label{fig3}
\end{figure}

Specifically, one has to consider the interdependence
of $m$ and $\tau$.
The estimation of $m$ depends on the selection of $\tau$ ($\rho$) 
as we show in Fig.\,\ref{fig3} for the Lorenz system. 
The proportion of false nearest neighbors does not fall to
zero for the same $m$ as $\tau$ increases but rather the estimated $m$ 
increases slowly with $\tau$. Thus, the estimation of $m$ is somewhat 
arbitrary unless the method finds the same $m$ for a sufficiently large 
range of $\tau$ values. For a very small $\tau$, there is a 
typical underestimation of $m$. Such a $\tau$ forces the attractor to 
lie near the diagonal in ${\mathbb{R}}^m$. Increasing $m$ by one 
has little effect on the geometry of the attractor as it will still lie 
near the diagonal of ${\mathbb{R}}^{m+1}$. All the points will apparently look 
as true neighbors leading to a wrong conclusion. 

The method is very sensitive to noise giving larger values of $m$ for 
noisy data as pointed in \cite{Aleksic91} and \cite{Abarbanel93b}.
In fact, the effect of noise is greater for larger values of $\tau$. 
This is a serious drawback of the method because in real applications 
we are led to choose a larger $m$ than we really need.  
This problem is particularly relevant for MOD, where the projections are 
chosen without regard to noise filtering which is partly accomplished 
using SSA-reconstructions \cite{Albano88}.
  
Another method that has been suggested to estimate $m$ is based on 
truncating the singular spectrum of SSA (for details see \cite{Broomhead86} 
and \cite{Vautard92}). In fact, the idea behind this 
linear approach is, given the hyperspace of dimension $p$, to 
find the smallest subspace (hyperplane) that approximately bounds the 
attractor. This subspace is spanned by the eigenvectors
corresponding to the largest eigenvalues of the sample covariance
matrix, i.e. the directions where the attractor has the largest variance.
However, a strange attractor lies on a manifold which occupies all 
directions in the embedded space (very much like noise) and a clear 
cut-off is not expected \cite{Mees87}. 
On the other hand, if this approach is implemented locally it can reveal the 
dimension of the tangent space to the manifold and the averaging over a 
grid of local regions can give a robust estimate of $m$ as shown 
in \cite{Passamante89}.
However, this estimate depends on the choice of the
dimension $p$ of the hyperspace, i.e. the time window length $\tau_w$. 

From these remarks we conclude that many of the existing
methods for estimating $\tau$ and $m$ are based on somewhat arbitrary 
criteria and do not always guarantee good reconstructions. The 
performance depends on the problem at hand.

\section{The time window length - $\tau_w$}
\label{tau_w}

When analysing a time series one typically begins with an initial
reconstruction, and implements a non-linear method to this and 
other modified reconstructions until a stable result is attained.
Here we concentrate on the time window length $\tau_w$ to determine 
the reconstruction.

There is probably no uniquely best way to choose an initial $\tau_w$. 
We will argue that it may be reasonable to set $\tau_w$ equal to the 
``memory'' of the system, i.e. the measurement record needed
to determine future observations as reliably as possible. 
For practical reasons, one would like the shortest possible $\tau_w$.
Geometrically, one could associate such a $\tau_w$ with the {\em mean
orbital period} $\tau_p$, i.e. the mean time between two consecutive visits
to a local neighborhood. For low-dimensional chaotic systems 
showing pseudo-periodicity, the mean orbital period could naturally 
be associated with the mean time between visiting a Poincare section.

For several chaotic systems, $\tau_p$ carries significant
information about the dynamics.
For systems that generate attractors with a sheet-like structure
in ${\mathbb{R}}^3$ (see for example \cite{Medio92}), it can be shown that the 
Poincare section gives points that in a scatter plot lie approximately 
on a curve, which is the one dimensional manifold that embeds an 
attractor very much like the strange attractor of the logistic map. 
The same result may be obtained by selecting the points from the extremes
or maxima of the time series directly instead of using reconstruction 
and Poincare section. This has been shown for the 
Lorenz system \cite{Ott93} and the R\"{o}ssler system 
\cite{Olsen85}. 
We found similar results studying the oscillations
of other systems with sheet-like structure, such as the 
Rabinovich-Fabrikant system \cite{Rabinovich79} and the Mackey Glass 
system for $\Delta=17$ \cite{Mackey77} (for details of this system see 
below). 

As indicated above, the procedure suggested here requires only an 
initial estimate of $\tau_w$ which is subsequently adjusted. 
Given only a set of observations, a very 
simple solution is to select the initial $\tau_w$ as the mean {\em 
time between peaks} ({\em tbp}) of the original time series. In general,
{\em tbp} will be less than $\tau_p$, and thus it is 
natural to consider {\em tbp} a lower limit. 
For a low dimensional 
system, e.g. defined asymptotically in ${\mathbb{R}}^3$, it is reasonable 
to assume that an orbital period corresponds to an oscillation when
projected down to the observed axis, and thus $\tau_p = tbp$.
For more complicated systems in higher dimensional spaces, a complete 
orbit may form more than one oscillation. In that case, $\tau_p$ should
be estimated as the average over a pattern of oscillations.

\begin{figure}[htb] 
\hspace{-5mm}
\centerline{\hbox{\psfig{file=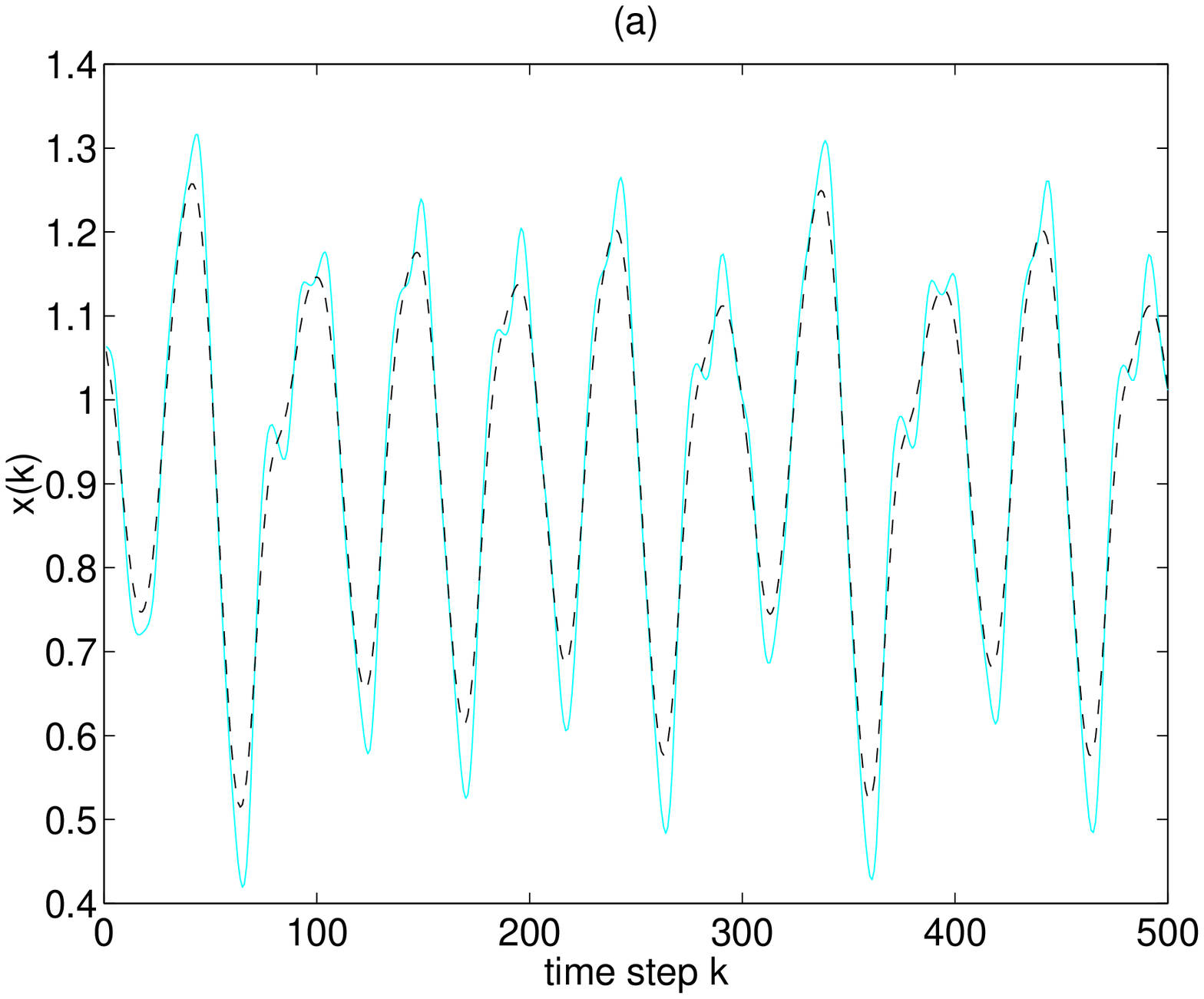,height=5cm,width=5cm}
                 \psfig{file=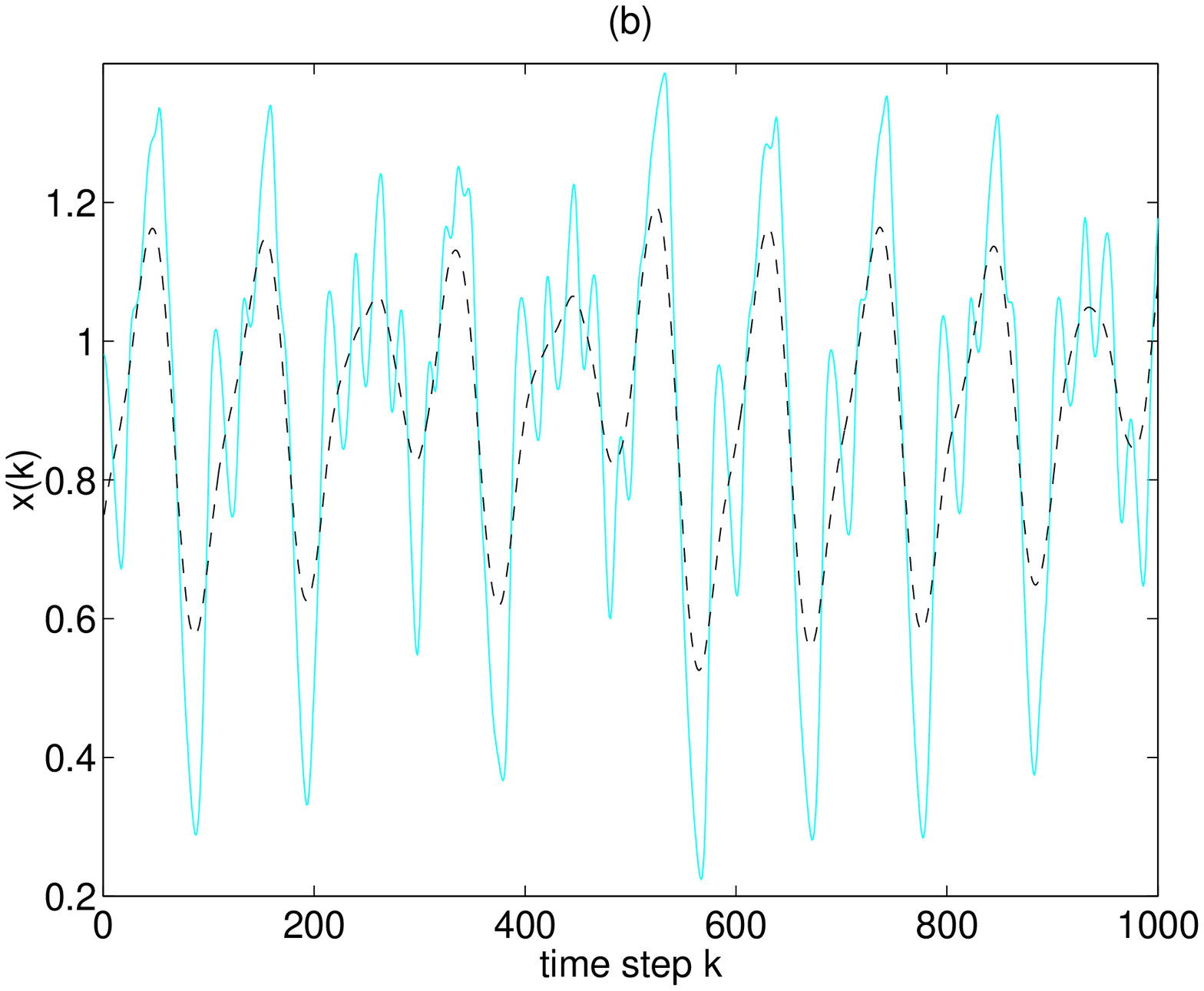,height=5cm,width=5cm}
                 \psfig{file=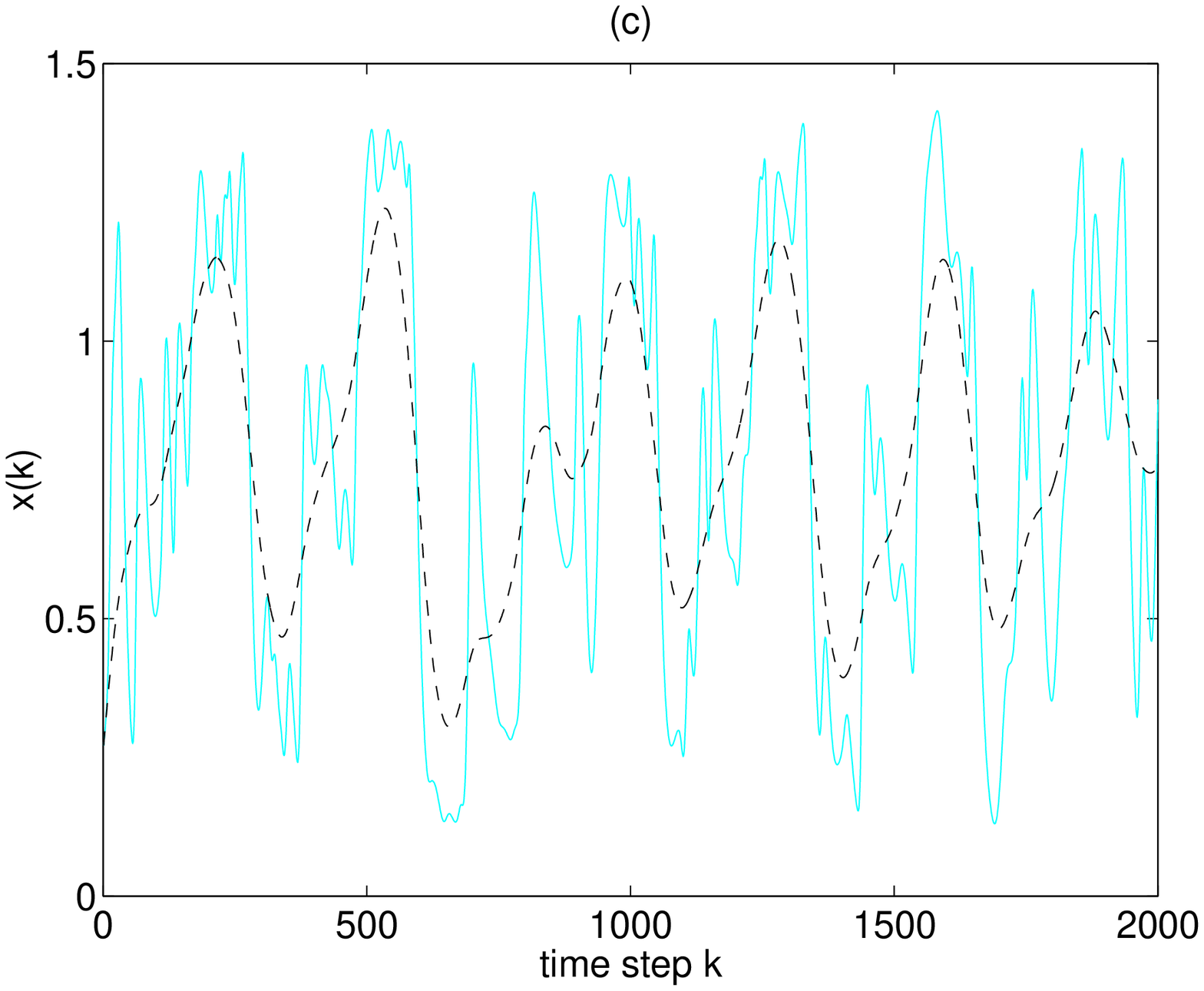,height=5cm,width=5cm}
           }}
\caption{The solid grey lines in all three figures are for segments
         of the Mackey Glass time series for different $\Delta$
         and the stippled lines after smoothing with a $k$-FIR filter.  
         (a) $\Delta = 17$, $\tau_s = 1$, and $k = 10$.
         (b) $\Delta = 30$, $\tau_s = 1$, and $k = 30$.
         (c) $\Delta = 100$, $\tau_s = 1$, and $k = 80$.}
\label{fig4}
\end{figure}

The equation of Mackey Glass \cite{Mackey77} 
\be
  \dot x = \frac{0.2x(t- \Delta)}{1+[x(t-\Delta)]^{10}} +0.1x(t) 
\ee
is a good example to show how one can find lower
limits for $\tau_w$ from the oscillations of the time series. 
This time delay differential equation was discretized following the 
iterative scheme in \cite{Ding93}, and 
segments of the time series for different $\Delta$ are shown 
in Fig.\ref{fig4} with solid grey lines.

For $\Delta = 17$, the attractor is low dimensional ($d \simeq 2$ 
\cite{Grassberger83a}) and an orbital period can be assumed to
correspond to a single oscillation (solid grey line in 
Fig.\,\ref{fig4}a).
Then $\tau_p$ can be easily estimated as $tbp$ after filtering the time
series to avoid close peaks that do not correspond to distinct 
oscillations (stippled black line in Fig.\,\ref{fig4}a), and thus 
for $\Delta = 17$ we can conclude that 
$\tau_w \geq \tau_p = tbp \simeq 50$ time units. 

For $\Delta = 30$, the attractor has a higher dimension ($d \simeq 3$
\cite{Grassberger83a}) 
and as Fig.\,\ref{fig4}b shows, in many parts of the time
series there are systematic variations 
over a pattern of oscillations (often comprised of a small and a 
large oscillation), approximately repeating itself. 
Filtering gives a new time series with one peak for 
each such pattern, facilitating the computation of $\tau_p$ from the $tbp$ of 
the filtered time series
giving $\tau_w \geq \tau_p \simeq 100$.

For $\Delta = 100$ in Fig.\,\ref{fig4}c, the attractor is 
much more complicated ($d \simeq 7.1$ \cite{Ding93}) and therefore it is 
difficult to observe patterns of oscillations that repeat themselves 
(but not as difficult as to make Poincare sections). 
However, in some particular 
parts of the time series, consecutive similar patterns may be observed 
showing implicit correspondence to orbital periods 
(see Fig.\,\ref{fig4}c). Hard
filtering allows us even to assign a peak to each pattern giving 
$\tau_w \geq \tau_p \simeq 330$.
Note that filtering is performed only in order to discern the representative
peaks, especially for higher dimensional systems. Noisy time series 
should be filtered anyway, before estimating $\tau_p$ to avoid
the fake peaks that are due to noise. 

\begin{figure}[htb] 
\hspace{-5mm}
\centerline{\hbox{\psfig{file=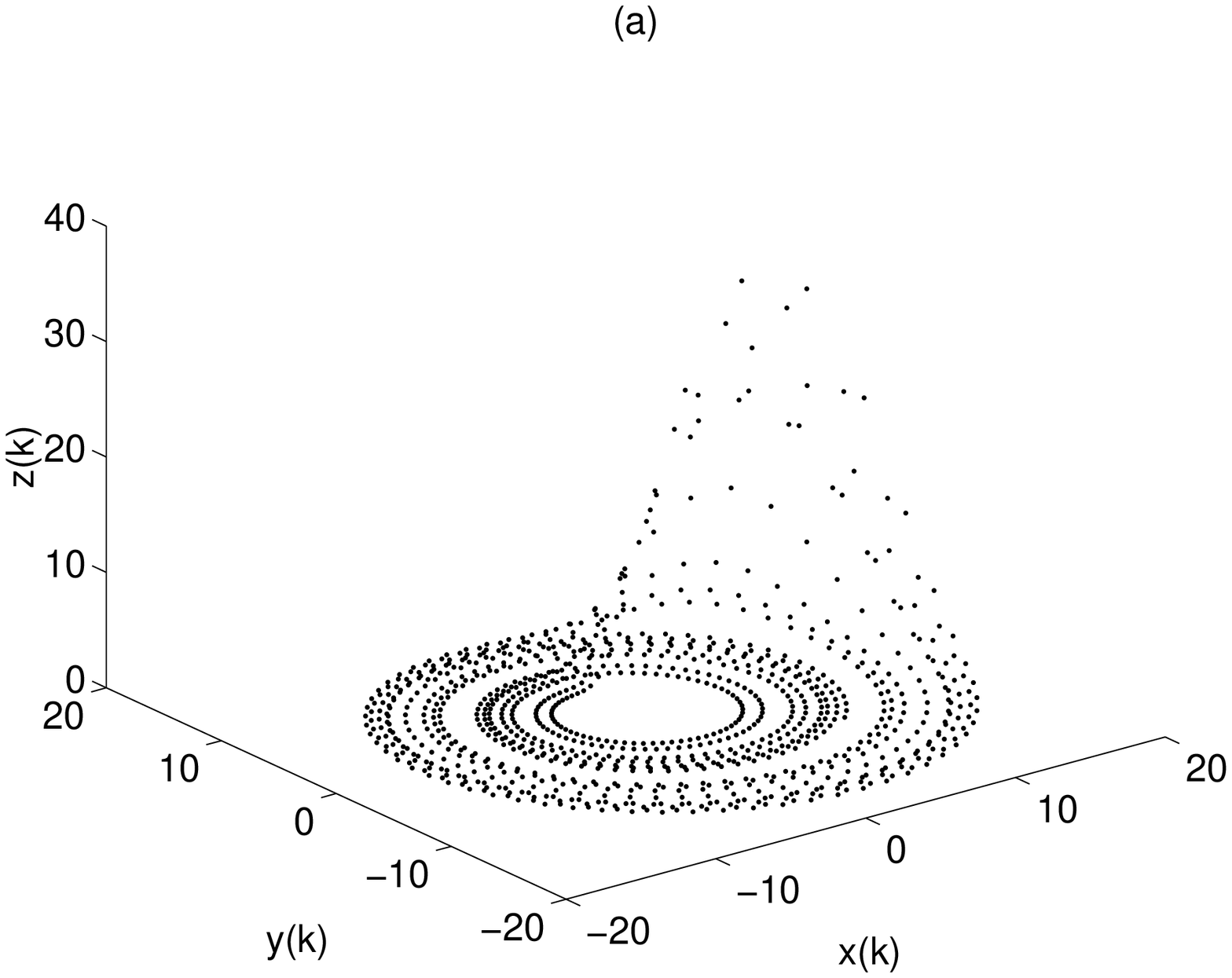,height=5cm,width=5cm}
                 \psfig{file=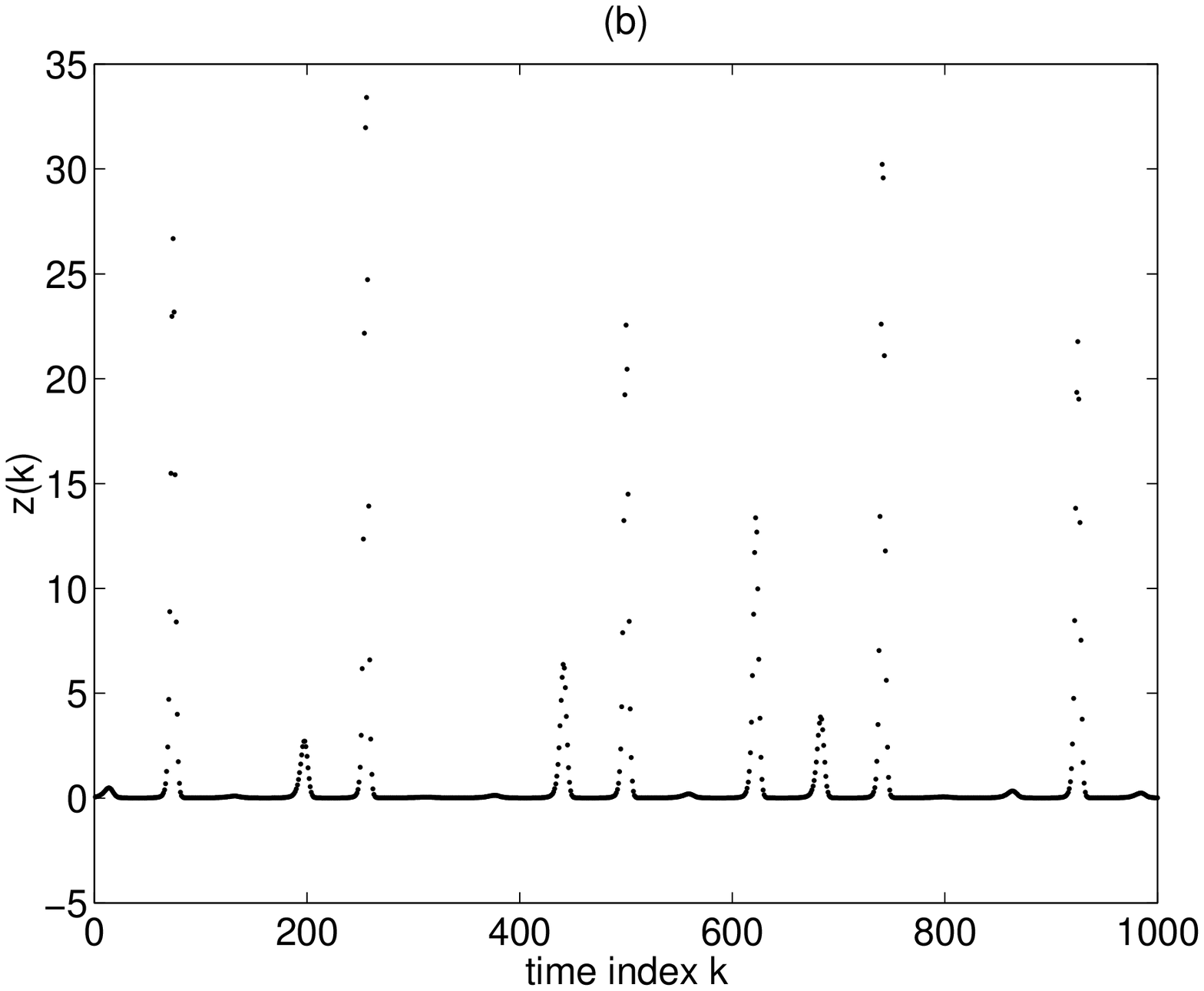,height=5cm,width=5cm}
                 \psfig{file=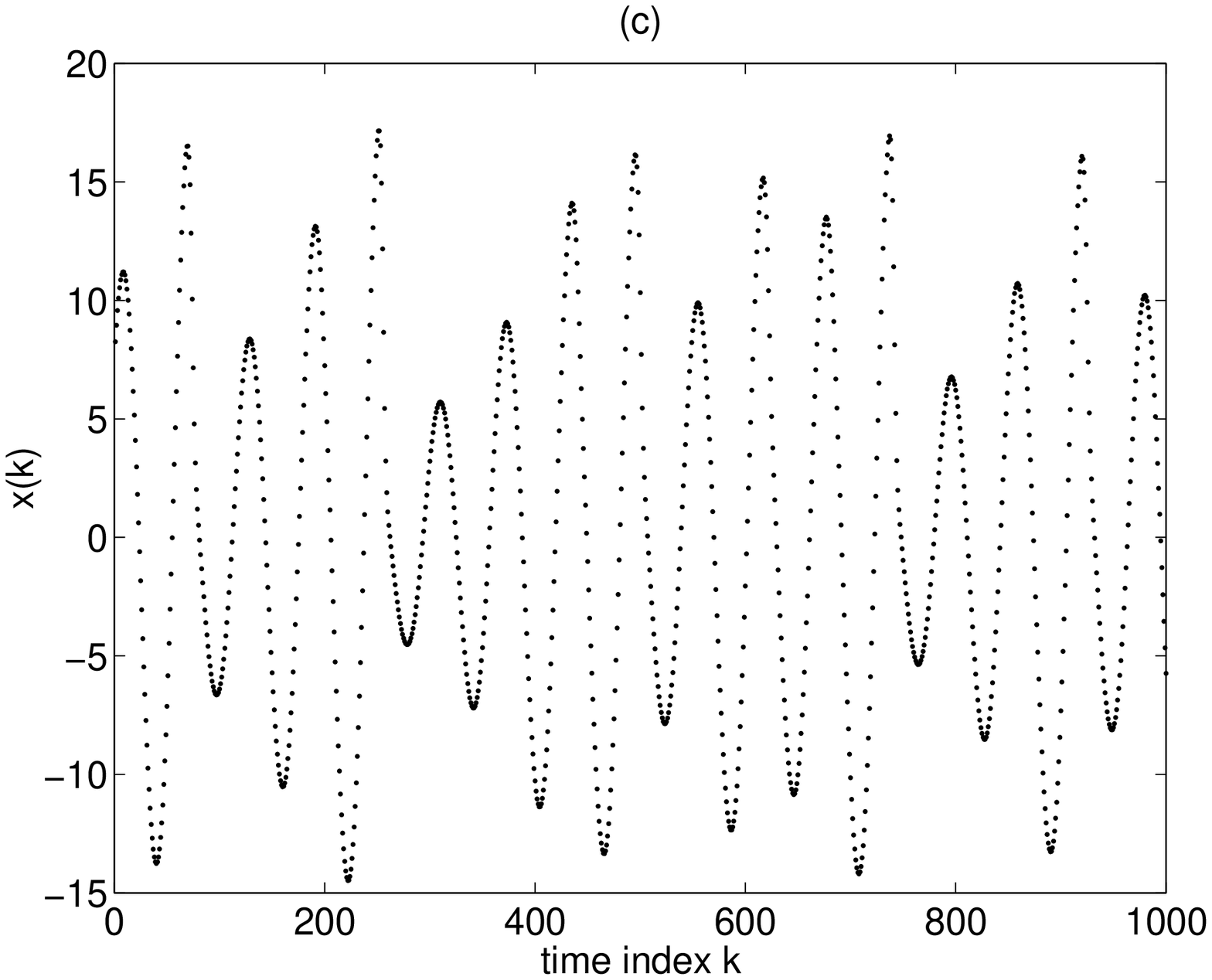,height=5cm,width=5cm}
           }}
\caption{(a) A trajectory of the R\"{o}ssler system in ${\mathbb{R}}^3$.
         (b) Measurements of the $z$ variable of the trajectory.
         (c) Measurements of the $x$ variable of the trajectory.
         Note that the oscillations of the time series in (b) do not
         reveal all orbital periods associated with the trajectory
         while in (c) they do.} 
\label{fig5}
\end{figure}

Up to this point we have assumed that the measurement function is 
well defined according to Takens' generic assumptions, so that the 
oscillations in the observed time series do reflect the periodic-like 
orbits of the original system and vice versa.
However, this is not always the case and as an example of a 
``good'' and ``bad'' mapping let us consider the $x$ and $z$ variable of 
the R\"{o}ssler system \cite{Roessler76} (see Fig.\,\ref{fig5}). 
In the time series of the $x$ variable, the oscillations represent 
the real orbits while in the time series of the $z$ variable the 
orbital periods can hardly be recognized. 
In the latter case, an analysis will fail to identify the correct 
attributes of the system unless a very large 
amount of data is provided to compensate for the bad mapping. 
We found, for example, that for measurements over the same epoch, the 
correlation dimension of the R\"{o}ssler attractor was well 
estimated by the $x$-measurements but significantly underestimated by the 
$z$-measurements due to the ``knee'' phenomenon we discuss below.

We here suggest working directly in the time domain to estimate $\tau_w$
instead of considering periods corresponding to dominant frequencies as
suggested by \cite{Broomhead86} and \cite{Grassberger91}.
Chaotic data will in general not show well defined frequency 
peaks. Other suggestions regarding $\tau_w$ have been presented in the 
literature \cite{Caputo86}, \cite{Gibson92} and 
\cite{Rosenstein93}. Some attempted to estimate $\tau_w$ based on
decorrelation criteria from the autocorrelation function and the 
mutual information \cite{Albano88}, \cite{Albano91} and 
\cite{Martinerie92}. In one paper treating this issue, \cite{Albano88}, 
lower and upper limits for $\tau_w$ where based 
on the autocorrelation function and it was proposed to set 
$\tau_c \leq \tau_w \leq 4\tau_c$, where $\tau_c$ is the correlation
time defined as the delay where the autocorrelation function is $1/e$. 
This lower limit is much smaller than $\tau_p$ for most systems.  
An upper limit for $\tau_w$ was given in \cite{Gibson92} by 
$2\sqrt{\frac{3\langle x^{(0)} \rangle}{\langle x^{(1)} \rangle}}$, where 
$\langle x^{(0)} \rangle$ and $\langle x^{(1)} \rangle$ are the mean 
values of the time series and its first derivative, respectively. We found 
that for many systems this upper limit is also smaller than $\tau_p$.

\section{Correlation dimension and $\tau_w$ }
\label{cordimtau_w}

We now discuss the use of $\tau_w$ in the time series analysis.
A natural procedure is to start with an initial $\tau_w$ 
and perform calculations -- in this case computing the correlation 
dimension $\nu$ -- for a sufficiently large $m$.
Then $\tau_w$ is modified, the calculations repeated, and so on.
To be able to conclude that a valid result has been 
obtained, reasonably stable values have to be found over a range
of $\tau_w$ values.

First we define the correlation integral $C(r)$, a statistic that 
measures the fraction of points on the attractor being less than $r$ 
units apart
\be
  C(r) =  \frac{1}{N(N-1)} 
          \sum_{i,j=1, |i - j| > K}^{N} \Theta(r - ||\mathbf x_i - \mathbf x_j||)
\ee 
where $\Theta(x)$ is the Heaviside function, defined as 
$\Theta(x)=1$ for $x \geq 0$ and $\Theta(x)=0$ for $x < 0$, and $K$
is used to omit time-correlated points in the computation of $C(r)$.
The Euclidean norm is used because it gives more robust 
results in the presence of noise \cite{Kugiumtzis95c}. 
For deterministic systems, the correlation integral scales as 
$C(r) \sim r^{\nu}$, where theoretically $r \rightarrow 0$.
Preferably, $\nu$ should be estimated from the slope of the
graph of $\log C(r)$ against $\log r$ over a sufficient
range $[r_1, r_2]$ of small interdistances.
However, due to noise or to limited data, an approximately constant 
slope may be maintained only for larger values of $r_1$ and $r_2$. 
We chose $r_2/r_1 = 4$ for the length 
of the interval, and searched over all such intervals  
to find the one where the computed $\nu$ varied least\footnote{To 
compute the slope 
for each $r$ we use the best fit slope for three values, the
current $r$, the previous and the next.}. The mean value of the slope
in this interval is the estimated $\nu$, and it is always reported 
together with the standard deviation (shown with bars in the 
following figures).

\begin{figure}[htb] 
\centerline{\hbox{\psfig{file=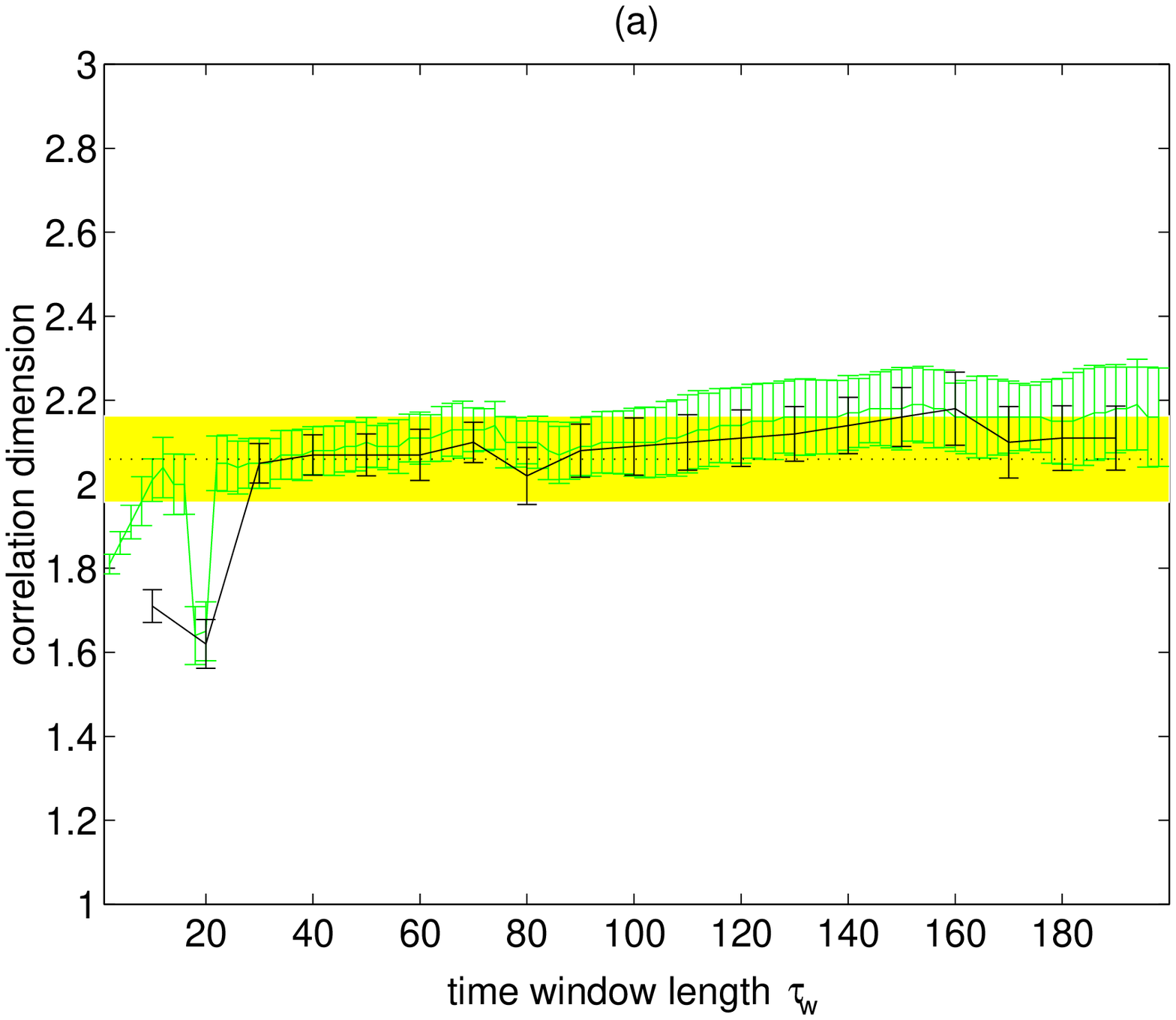,height=5cm,width=5cm}
                 \psfig{file=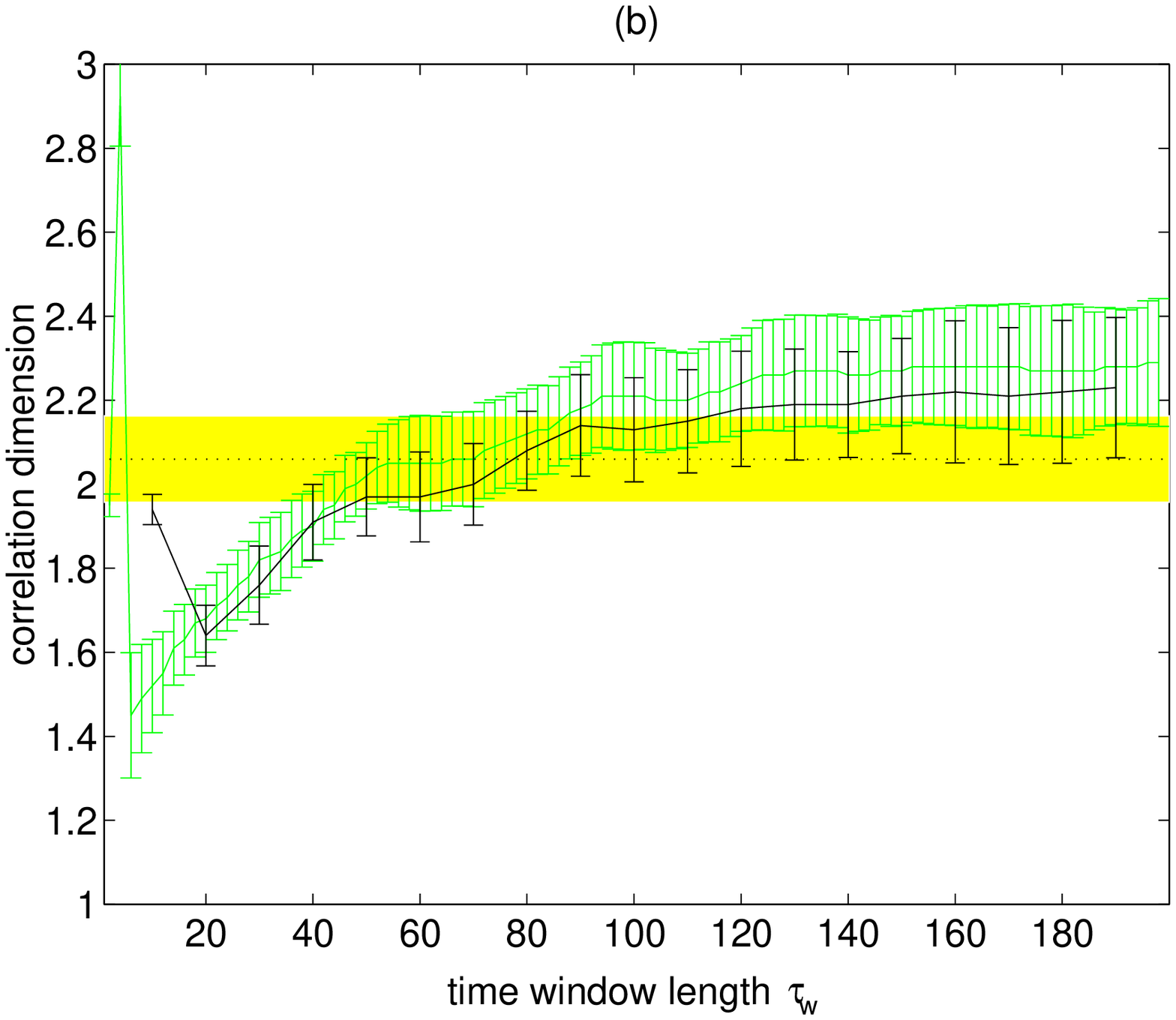,height=5cm,width=5cm}
           }}
\caption{Plot of the correlation dimension estimate $\nu$ 
         for MOD reconstruction with different $\tau_w$ 
         for time series of the $x$ variable of the Lorenz system.
         The bars denote the standard deviation of the estimate.
         In each figure the grey curve with grey error bars correspond to 
         $\rho = 2$ while the black ones to $\rho = 10$.
         In (a) the estimation is based on the clean time series of 
         4000 data sampled with $\tau_s=0.02$ and in (b) on the same 
         data but corrupted with $5\%$ noise. The horizontal stippled line 
         shows the correct plateau for $\nu = 2.06$ and the shaded area
         the confidence interval of $\pm 5\%$ of the correct $\nu$.} 
\label{fig6}
\end{figure}

A key observation is that the estimate of the correlation dimension 
of a chaotic time series (clean or noisy)
is approximately the same under variations of the parameters $\rho$ 
and $m$ while keeping $\tau_w = (m - 1) \rho$ fixed (assuming that 
$m$ is always larger than the dimension of the attractor). Only 
few workers seem to have thought along these lines 
(\cite{Caputo86}, \cite{Albano88}, \cite{Grassberger91} and 
\cite{Rosenstein93}). The typical features are  demonstrated in 
Fig.\,\ref{fig6} which shows the correlation dimension 
estimates for different $\tau_w$
for clean and noisy data from the Lorenz system.
Note how the grey and black curves match for the clean data in
Fig.\,\ref{fig6}a. They correspond to the same $\tau_w$ but with 
$\rho = 2$ and $\rho = 10$, respectively. 
Once $\tau_w$, and thus the $p$-dimensional hyperspace, has been 
determined, the particular projection chosen is not critical
as long as the projection is sufficient, i.e. $m > \nu$ and 
$\rho \simeq \frac{p-1}{m-1} \equiv \frac{\tau_w}{m-1}$. 
This is so, 
because the interdistances of points remain statistically the same in 
${\mathbb{R}}^p$ and in ${\mathbb{R}}^m$. Considering all the coordinates or 
only the selected subset has the same effect on the computation of the 
interdistance as long as a suitable norm is used, e.g. the Euclidean 
norm \cite{Kugiumtzis95c}.

When white noise is added to the clean 
Lorenz data (Fig.\,\ref{fig6}b) the two curves still match 
but now show an increasing trend with $\tau_w$.
The estimation of $\nu$ is more sensitive to the choice 
of $\tau_w$ in the presence of noise.

Results for the estimation of $\nu$ from noisy data or few data
(compared to the minimum number of data required) should be 
interpreted with caution because they are derived from scaling 
properties based on large $r$. For smaller $r$, the 
scaling is corrupted by noise or distorted due to few neighbors
in state space. 
In the case of attractors with different scaling properties for small 
and large $r$ (a phenomenon referred to as a ``knee'' \cite{Theiler90}), 
erronous estimates are obtained from the scaling for large $r$ when 
noise or insufficient data length mask the correct scaling for small 
$r$. Such a phenomenon is observed for the $z$-measurements of 
the R\"{o}ssler system mentioned before. The correct scaling 
($\nu \simeq 2.01$) 
can be only detected for very small inter-point distances $r$ 
requiring a very large number of data, otherwise another scaling 
is detected for larger $r$, underestimating $\nu$.   

\begin{figure}[htb] 
\centerline{\psfig{file=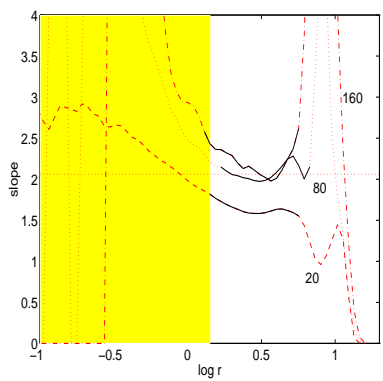,height=5cm,width=5cm}}
\caption{Plot of the slope of the graph $\log C(r)$ against $\log r$
         for the time series of 4000 data from the Lorenz system,
         sampled with $\tau_s = 0.02$ and with $5\%$ additive noise. 
         The three curves are derived from reconstructions with 
         $\rho = 10$ and $m=3$ (minimum embedding dimension), $m=9$
         and $m=17$ and are identified by the length of $\tau_w$ marked
         on the figure. The scaling interval of least variation is 
         denoted with the black solid line segment for each slope
         curve. The grey area shows roughly the region where inter-point 
         distances are corrupted by noise leaving a small
         interval of $r$ to estimate $\nu$ and making the choice of
         $\tau_w$ critical. The horizontal stippled line shows the
         correct plateau for $\nu \simeq 2.06$.}
\label{fig7}
\end{figure}

The estimation of $\nu$, even when it is constrained only to large $r$, 
is not straightforward as it varies with $\tau_w$ and a typical 
situation is shown in Fig.\,\ref{fig7} for Lorenz system.
Too small $\tau_w$ ($\tau_w = 20$) or too large $\tau_w$ ($\tau_w = 160$) 
gives uncertain and wrong estimates while for $\tau_w$ larger than but 
still close to $\tau_p = 50$ (here $\tau_w = 80$)\footnote{For this
time series the periods of the oscillations vary a lot and thus the
estimate $\tau_p$ has large variance and does not completely 
indicate the ``memory of the system''.} 
the scaling is clear indicating a reliable estimate.
On the other hand, the range of suitable $\tau_w$ depends on the 
length of the time series; the longer the time series, the broader 
the limits for $\tau_w$. 
Noise also restricts $\tau_w$ from above because the slope 
curves derived for increasing $\tau_w$ do not saturate. 
Setting a criterion for the acceptance of the $\nu$-estimate,
e.g. $\pm 5\%$ of the correct value, an upper limit $\tau_n$ 
for the range of $\tau_w$ may be found which varies with the 
amplitude of the noise (e.g. $\tau_n \simeq 110$ for 
Fig.\,\ref{fig6}b). It is thus expected that the scaling gets 
distorted as $\tau_w$ icreases over $\tau_n$ giving less 
confident estimates as shown with the slope curve for $\tau_w = 160$ 
in Fig.\,\ref{fig7}. So, when the time series is corrupted 
with noise, the $\nu$-estimates are more biased and the 
interval $[\tau_p, \tau_n]$ of the accepted $\tau_w$ 
shrinks from above, and it may be no reliable estimate of $\nu$ 
for any $\tau_w$ if the impact of noise is so large 
that $\tau_n$ decreases to the level of $\tau_p$. 

Thus when estimating $\nu$ from a limited number of noisy 
data we seek the range of $\tau_w$ that gives clear 
scaling for large $r$ keeping in mind that the results are still 
ambiguous due to the possible different scaling for small and
inaccessible $r$ (the ``knee'' phenomenon). In the sequel, 
we consider in more detail 
simulated data corrupted with noise as well as real data.

\subsection{Noisy synthetic data}
\label{simreal}

Most of the time series we use here have length $N=4000$ adjusting 
the sampling time $\tau_s$ 
accordingly in order to have enough oscillations as well as enough 
samples for each oscillation. It follows that the number of data 
points is not the best measure of the record length. We therefore also
quote the number of $\tau_p$ within the record, denoted $\#\tau_p$, 
together with the number of samples in $\tau_p$.
Note that under changes of the reconstruction parameters or the noise 
amplitudes, the values $r_1$ and $r_2$ of the scaling interval $[r_1, r_2]$ 
that gives $\nu$-estimates with least variance may change as well. 

Results for the time series from the $x$-variable of the Lorenz system 
with $\tau_s = 0.02$ and $\tau_p \simeq 50$ and $\#\tau_p \simeq 80$ 
were shown in Fig.\,\ref{fig6}. For the clean data, legitimate 
estimates of $\nu$ (within $\pm 5\%$ of the 
correct $\nu = 2.06$ shown as a shaded zone in the figure) were 
obtained for a large interval of $\tau_w$ values beginning even 
lower than $\tau_p$. As $\tau_w$ is increased long
beyond $\tau_p$ the estimates increase somewhat and have larger variance. 
When $5\%$ white Gaussian noise is added to these data, 
the correlation dimension is underestimated significantly 
for $\tau_w < \tau_p$, and for $\tau_w > \tau_n \simeq 110$, $\nu$ is 
overestimated with larger variance.

\begin{figure}[htb] 
\centerline{\psfig{file=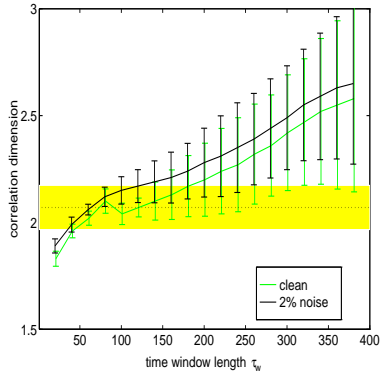,height=5cm,width=5cm}}
\caption{Plot of the correlation dimension estimate $\nu$ 
         for MOD reconstruction with different $\tau_w$ 
         for time series of the $x$ variable of the R\"{o}ssler system.
         The grey curve with grey error bars correspond to 
         the clean data and the black to the same data corrupted
         with $2\%$ noise. Here, $N=4000$, $\tau_s = 0.1$ and $\rho = 20$. 
         The horizontal stippled line shows the correct plateau for 
         $\nu = 2.01$ and the shaded area the confidence interval 
         of $\pm 5\%$ of the correct $\nu$.}
\label{fig8}
\end{figure}

The attractor derived from the $x$-variable of the R\"{o}ssler system 
has a simpler structure than the Lorenz attractor and about 
the same dimension. However, estimates of $\nu$ are
more dependent on the reconstruction parameters and the amplitude of the 
noise. The time series is sampled with $\tau_s = 0.1$ that 
gives $60$ samples in each oscillation and about $66$ oscillations, 
which are comparable to the $\tau_p$ and $\#\tau_p$ for the 
Lorenz data. In Fig.\,\ref{fig8}, the $\nu$-estimates are plotted against 
$\tau_w$ for the clean and noisy R\"{o}ssler data displayed with grey
and black error bars respectively, together with the $\pm 5\%$-zone of the
accepted range of $\nu$. Here, as well as in the following estimations, 
we keep $\rho$ fixed ($\rho = 20$ in Fig.\,\ref{fig8}) and vary $m$. 
This is done for convenience 
since the results are essentially the same for other combinations of $\rho$
and $m$ (refer back to Fig.\,\ref{fig6}).   
For the clean data, reasonable and confident $\nu$-estimates can be found 
for a small range of $\tau_p = 60 \leq \tau_w < 140$ (the grey error bars 
in the $\pm 5\%$-zone in the figure). When just $2\%$ white 
noise is added to these data, the small horizontal plateau seems to 
disappear (the black line in Fig.\,\ref{fig8}) and only 
$\nu$-estimates close to $\tau_p$ and above can be accepted, which is
in accordance with the proposed $\tau_w$.

\begin{figure}[htb] 
\centerline{\psfig{file=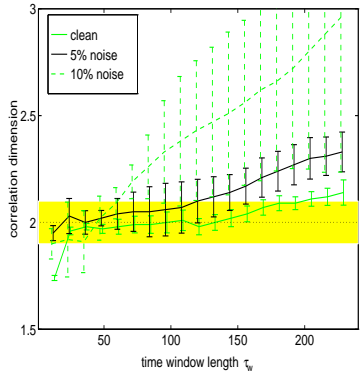,height=5cm,width=5cm}}
\caption{Plot of the correlation dimension estimate $\nu$ 
         for MOD reconstruction with different $\tau_w$ 
         for time series of the Mackey Glass equation for $\Delta = 17$.
         The solid grey curve with solid grey error bars correspond to 
         the clean data, the solid black to the noisy data with $5\%$
         noise and the stippled grey to the noisy data with $10\%$ noise. 
         Here, $N=4000$, $\tau_s = 1$ and $\rho = 12$. 
         The horizontal stippled line shows the correct plateau for 
         $\nu = 2$ and the shaded area the confidence interval 
         of $\pm 5\%$ of the correct $\nu$.}
\label{fig9}
\end{figure}

The Mackey Glass attractor for $\Delta = 17$ has the same dimensionality
as the two last attractors but gives less biased estimates of $\nu$.
For $\tau_s=1$, we found $\tau_p = 50$ and $\#\tau_p \simeq 80$ from single
oscillations. In Fig.\,\ref{fig9}, results from the estimation of $\nu$
are presented in the same way as for the R\"{o}ssler data.  
For the clean data, a very reliable $\nu$-estimate is derived over a 
large interval of $\tau_w$, $[20, 160]$ (from the 
$\pm 5\%$-criterion). When $5\%$ noise is added, 
confident estimates are obtained only close to $\tau_p$, 
and when $10\%$ noise is added, reasonable estimates are only 
obtained for $\tau_w \simeq \tau_p$. 

\begin{figure}[htb] 
\centerline{\hbox{\psfig{file=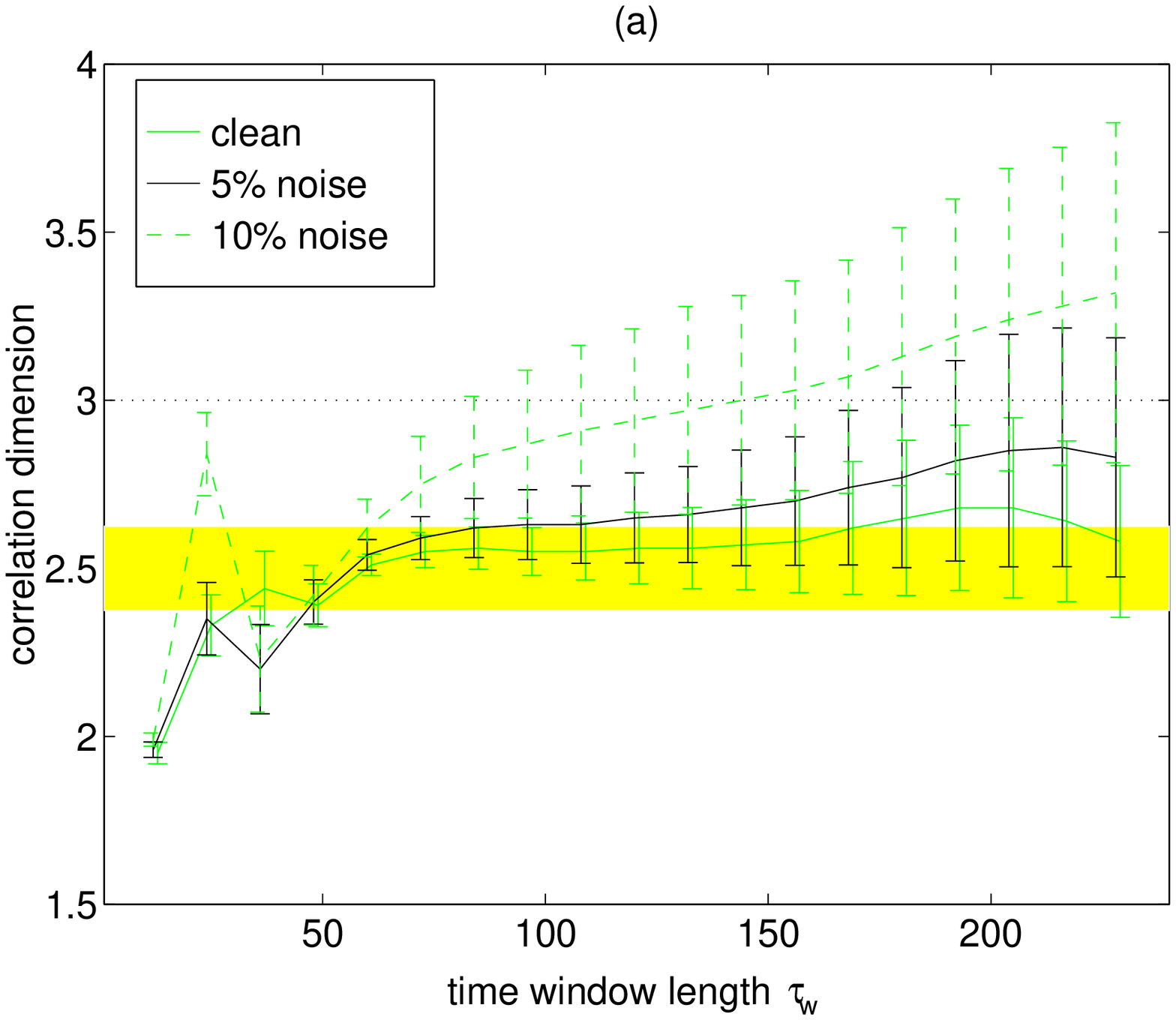,height=5cm,width=5cm}
                 \psfig{file=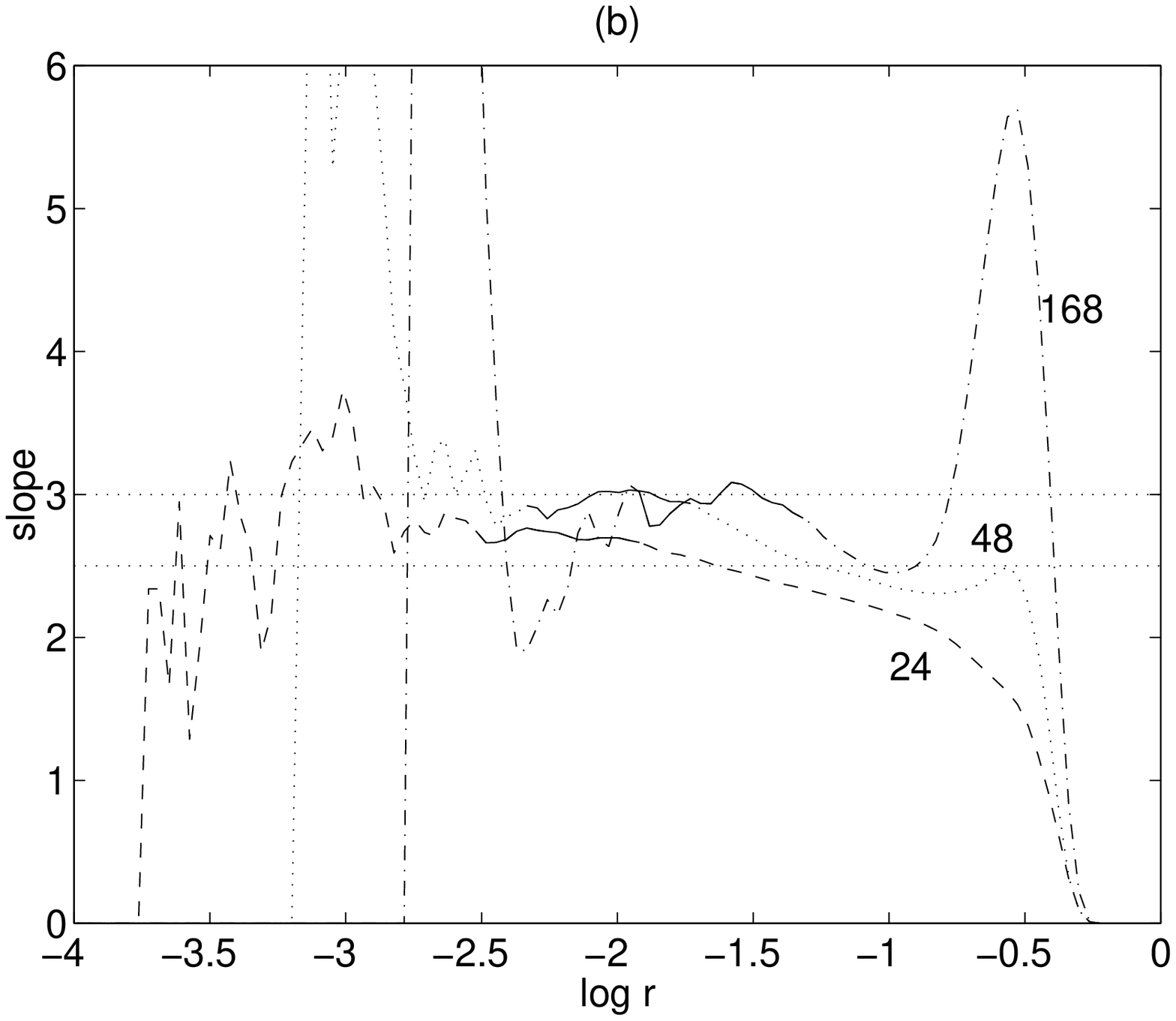,height=5cm,width=5cm}
           }}
\caption{(a) Plot of the correlation dimension estimate $\nu$ 
         for MOD reconstruction with different $\tau_w$ 
         for time series of the Mackey Glass equation for $\Delta = 30$.
         The solid grey curve with solid grey error bars correspond to 
         the clean data, the solid black to the noisy data with $5\%$
         noise and the stippled grey to the noisy data with $10\%$ noise. 
         Here, $N=4000$, $\tau_s = 2$ and $\rho = 12$. 
         The horizontal stippled line shows the correct plateau for 
         $\nu \simeq 3.0$ and the shaded area the confidence interval 
         of $\pm 5\%$ of the underestimated $\nu \simeq 2.5$.
         (b) Plot of the slope of the graph $\log C(r)$ against 
         $\log r$ for the same type of data but for $N=30000$. 
         The three curves are derived from reconstructions with 
         $\rho = 12$ and $m=3$, $m=6$ and $m=17$ and are identified 
         by the length of $\tau_w$ marked on the figure. The scaling 
         interval of least variation is denoted with the black solid 
         line segment for each slope curve. The two horizontal 
         stippled lines show the two scalings of this attractor.} 
\label{fig10}
\end{figure}

When $\Delta = 30$, the dimension of the attractor increases
to $\nu \simeq 3$ \cite{Grassberger83a}. However, using $N=4000$ and 
$\tau_s = 2$ an underestimate ($\nu \simeq 2.5$) was found.
For this $\tau_s$, the $\tau_p$ estimated with the mean time for 
patterns of two oscillations (cf. section\,\ref{tau_w}) is kept down 
to $\tau_p = 50$ and $\#\tau_p \simeq 80$, as for $\Delta = 17$. 
The results from estimation of $\nu$ for clean and
noisy data with $5\%$ and $10\%$ noise (shown in Fig.\,\ref{fig10}a)
assert the use of $\tau_p$ as a lower limit for $\tau_w$ and the decrease
of the interval of accepted values for $\tau_w$ from above and towards
$\tau_p$ as the amplitude of the added noise is increased.
The underestimation of $\nu$ is due to the limited number of data.
This attractor shows a ``knee'' structure, i.e. it has also another
scaling (the correct $\nu \simeq 3.0$) for small $r$ which can be 
detected only when many data are accumulated as shown in 
Fig.\,\ref{fig10}b. 
The slope for too small $\tau_w$ ($\tau_w=24$) underestimates
$\nu$ while for $\tau_w \geq \tau_p$ the correct scaling is achieved
(shown with the two curves for $\tau_w=48$ and $\tau_w=168$ in the 
figure). Note that these curves form a second scaling for larger $r$.

\begin{figure}[htb] 
\centerline{\psfig{file=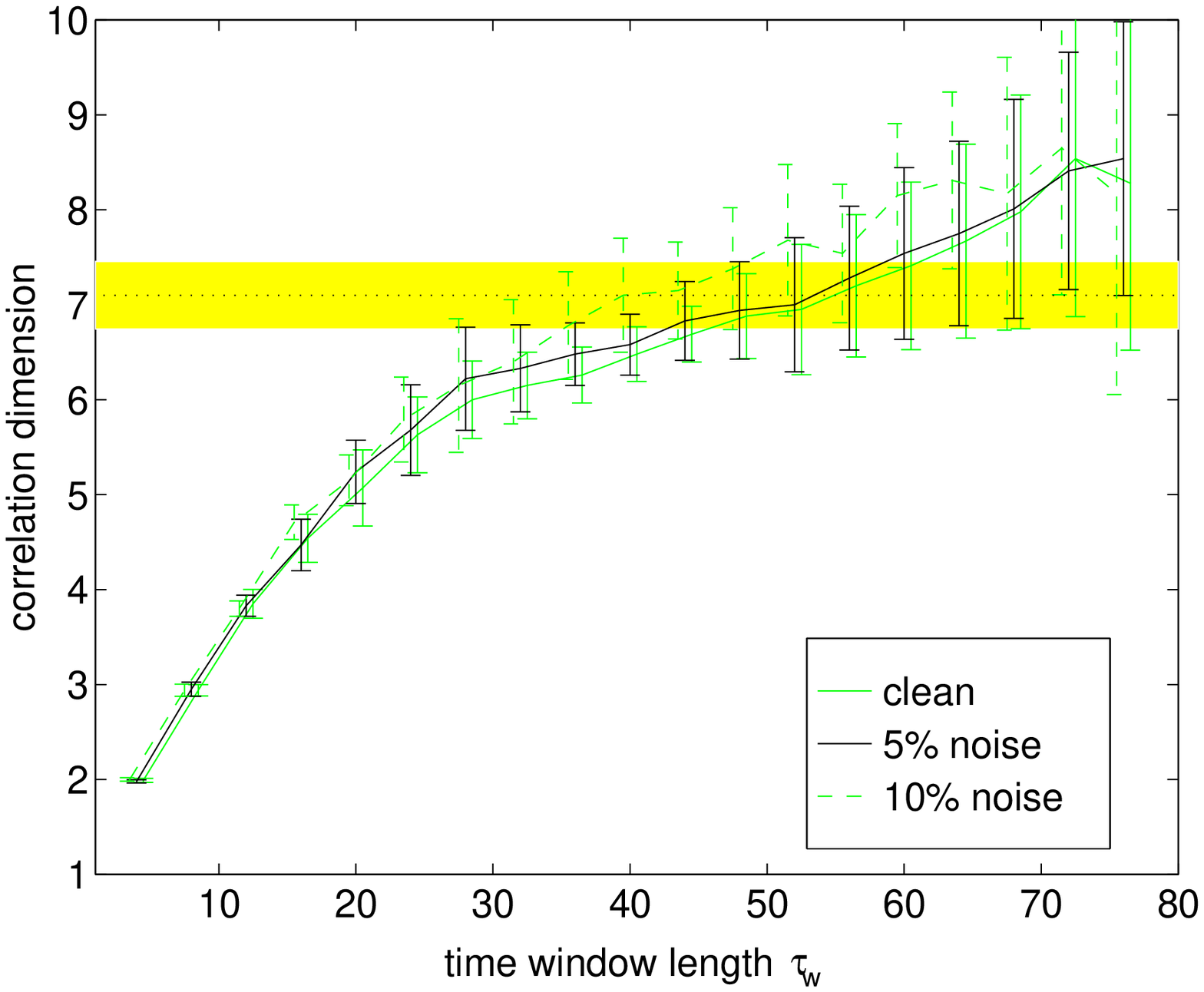,height=5cm,width=5cm}}
\caption{Plot of the correlation dimension estimate $\nu$ 
         for MOD reconstruction with different $\tau_w$ 
         for time series of the Mackey Glass equation for $\Delta = 100$.
         The solid grey curve with solid grey error bars correspond to 
         the clean data, the solid black to the noisy data with $5\%$
         noise and the stippled grey to the noisy data with $10\%$ noise. 
         Here, $N=4000$, $\tau_s = 10$ and $\rho = 4$. 
         The horizontal stippled line shows the correct plateau for 
         $\nu \simeq 7.1$ and the shaded area the confidence interval 
         of $\pm 5\%$ of the correct $\nu$.}
\label{fig11}
\end{figure}

For $\Delta = 100$, the Mackey Glass attractor gets high
dimensional with $\nu \simeq 7$ \cite{Ding93}. Our results show a 
slightly lower $\nu$ with as few as $N=4000$. We sampled the discretized
system with $\tau_s = 10$ in order to have enough, but not too many,
samples within the estimated mean orbital period, $\tau_p \simeq 33$, 
giving as many as $\#\tau_p \simeq 120$ 
repititions of the oscillation pattern that is assumed to correspond 
to an orbit of the underlying system.  
We deliberately keep the data record down to $N=4000$ in order to test 
our procedure for short time series (compared to the high dimensionality
of the system). The estimated $\nu$ is an increasing 
function of $\tau_w$ both in values and uncertainty, showing some stability 
in value and in variance for $\tau_p \simeq 30 \leq \tau_w \leq 45$. 
This is, however, an underestimation of $\nu$, possibly due to insufficient
data (see Fig.\,\ref{fig11}). 
Adding $5\%$ noise does not alter the $\nu$-estimates but just increases 
moderately the uncertainty of the estimates; when $10\%$ noise is 
added, the $\nu$-estimates for $\tau_w > \tau_p$ vary significantly from 
those of the clean data.  

These findings, as well as results for the Rabinovich-Fabrikant system 
\cite{Rabinovich79}, and the four-dimensional R\"{o}ssler Hyperchaos 
system \cite{Roessler79}, not shown here, confirm our suggestion for
estimating $\tau_w$ with $\tau_p$ giving the best estimates of $\nu$.
If the effect of noise or limited length of the time series is such that 
estimation of $\nu$ can be made only for a short range of $\tau_w$ values,
this is close to and little larger than $\tau_p$.

\subsection{Real data}

\begin{figure}[htb] 
\centerline{\psfig{file=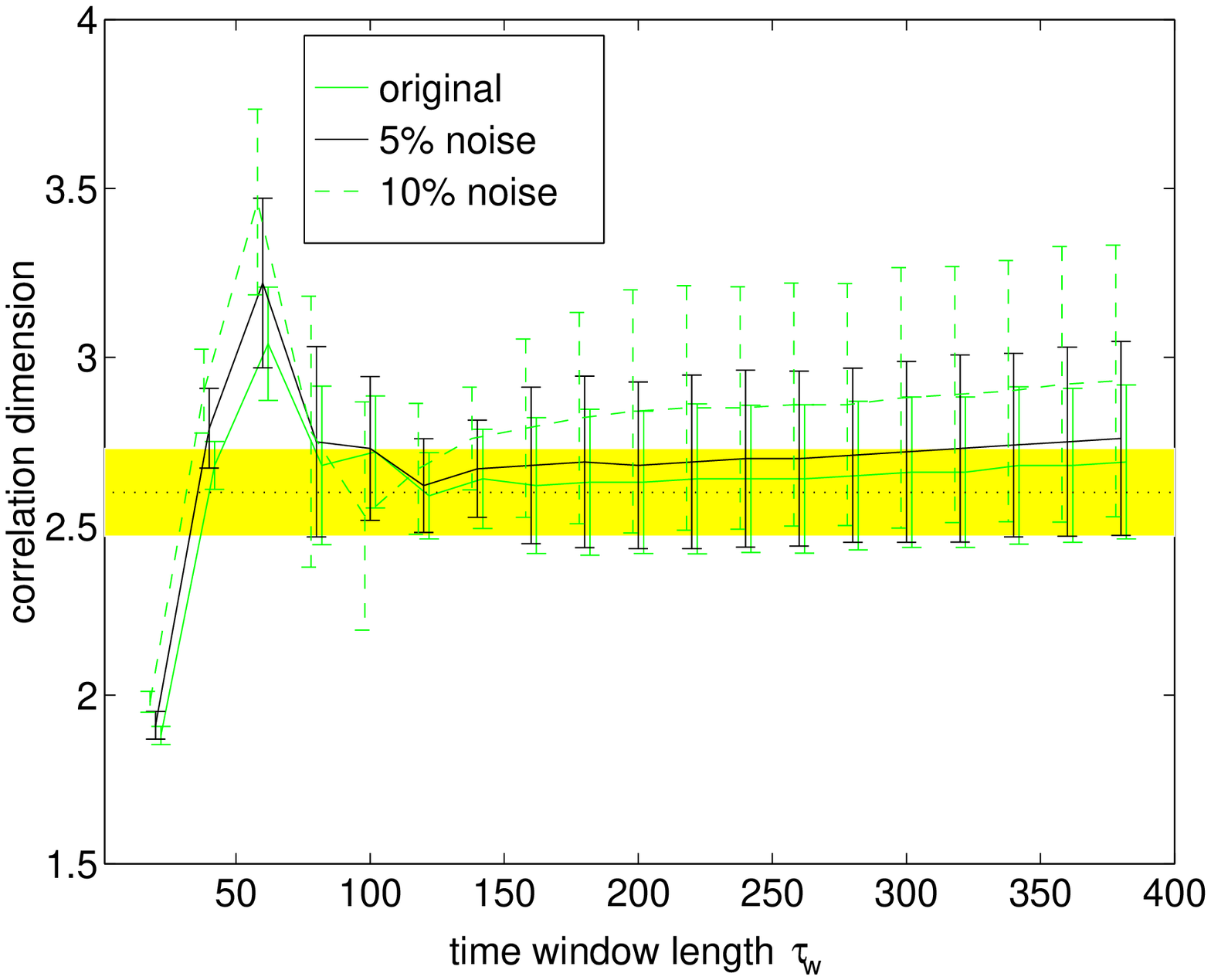,height=5cm,width=5cm}}
\caption{Plot of the correlation dimension estimate $\nu$ 
         for MOD reconstruction with different $\tau_w$ 
         for time series from the Taylor Couette experiment in the 
         chaotic regime. The solid grey curve with solid grey error bars 
         correspond to the original data, the solid black to the original
         data corrupted with $5\%$
         noise and the stippled grey to the original data corrupted with 
         $10\%$ noise. 
         Here, $\rho = 20$ is chosen for reconstructions varying with $m$.
         The horizontal stippled line shows the correct plateau for 
         $\nu \simeq 2.6$ and the shaded area the confidence interval 
         of $\pm 5\%$ of the correct $\nu$.}
\label{fig12}
\end{figure}

In addition to simulated data, observations from physical controlled 
experiments on low dimensional deterministic processes
should be used to assess the validity of non-linear methods. The
noise level is often insignificant in such cases. Here we use 
a time series of $N=4000$ samples from the Taylor Couette 
experiment in the chaotic regime. 
We estimated $\tau_p \simeq 75$ and $\#\tau_p \simeq 54$, but 
the results for the estimation of $\nu$ do not change for longer 
time records covering 
more oscillations (increasing either $N$ or $\tau_s$ if we insist on 
keeping $N$ small). 
Contrary to most of the previous results from simulated data
with noise, the estimated $\nu$ 
varies little with $\tau_w$ as shown in Fig.\,\ref{fig12}.
For all $\tau_w > \tau_p$ the etsimates are more or less fixed to 
$\nu \simeq 2.6$, 
approximately the value given in the literature \cite{Brandstater87},
with a slowly increasing uncertainty for $\tau_w > 150$. 
This indicates that there is little noise in the data and the dimension of
the chaotic attractor can be identified even with large $\tau_w$ (up to
$2 \tau_p$), so that the choice of $\tau_w$ is not critical. 
However, when we add noise to these data, to simulate a larger 
experimental uncertainty, the estimates have as expected a larger variance, 
but for $\tau_w$ close to $\tau_p$ the estimates are the same as for the 
original time series. For larger $\tau_w$ there is a systematic 
overestimation of $\nu$, showing again that the optimal $\tau_w$ for correct 
estimation is close to $\tau_p$.

We now turn to observational data that are not output of a controlled 
experiment, and concentrate on physiological data of the 
Electroencephalogram (fig13) from epileptic patients 
(e.g. see \cite{Jansen91}). 
Dimension estimation of physiological data has been a hot subject 
the last years. However, the results to date are not promising, partly because
different procedures are often used giving 
different $\nu$-estimates for the same type of data, and partly 
because these data do not seem to share the
same nice chaotic properties as the well-studied simulated 
data \cite{Kantz95}. 
Previous work on $\nu$-estimation of EEG epileptic signals reported low 
dimensional attractors of varying dimension between 2 and 6, according to 
the physiological nature of the data, the data acquisition process, the
computational scheme of estimation, as well as the parameter setting
for reconstruction (\cite{Rapp85}, \cite{Babloyantz86}, 
\cite{Frank90} and \cite{Pijn91}). 

\begin{figure}[htb] 
\centerline{\hbox{\psfig{file=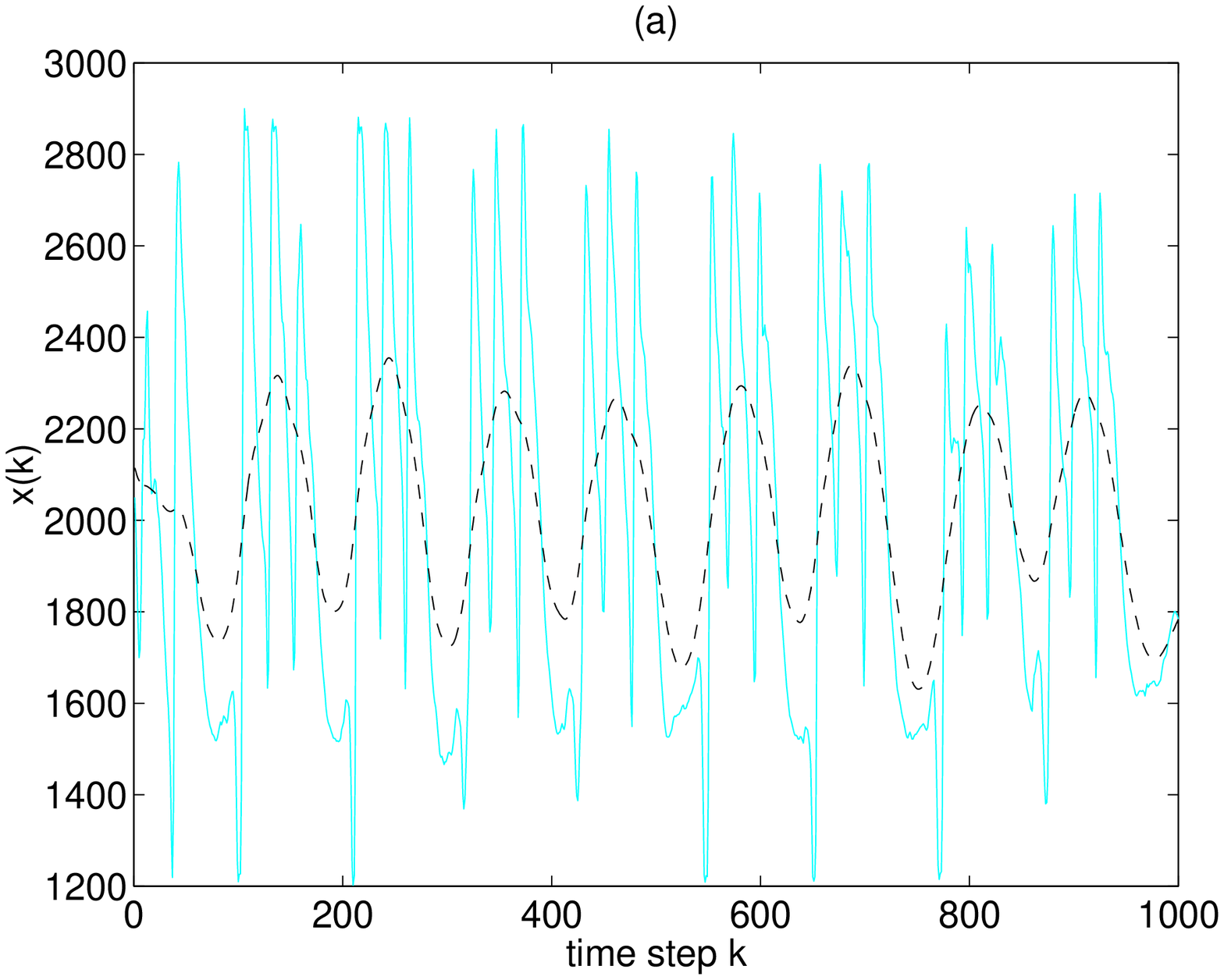,height=5cm,width=5cm}
                  \psfig{file=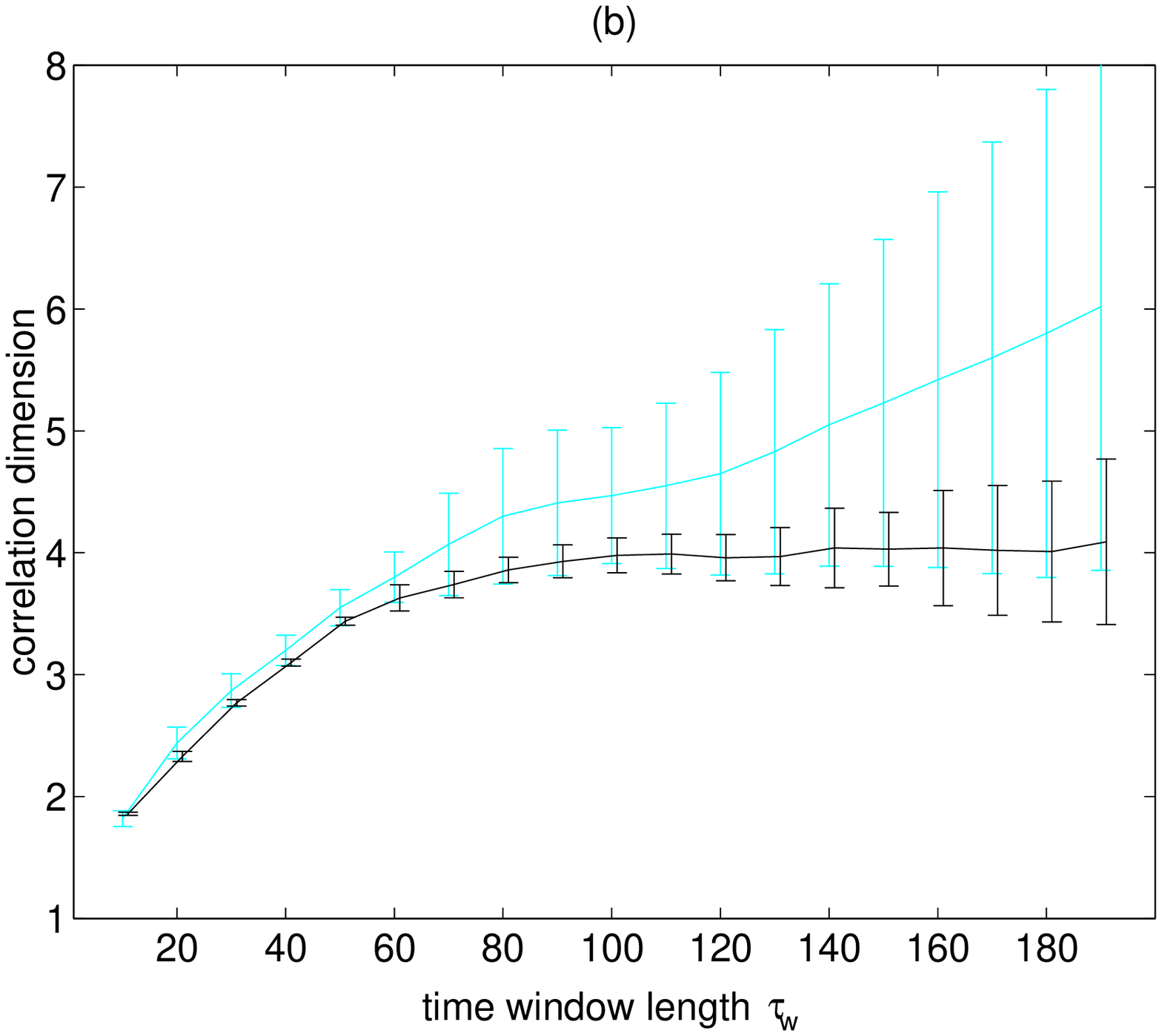,height=5cm,width=5cm}
           }}
\caption{(a) Segment of the EEG time series of an epileptic seizure 
         sampled with $\tau_s = 0.005sec$ (solid grey line) and after smoothing
         with a 40 point FIR filter (stippled black line).  
         (b) Plot of the correlation dimension estimate $\nu$ 
         for MOD reconstruction with different $\tau_w$ 
         for EEG time series in epileptic seizure.
         The grey curve with grey error bars correspond to 
         estimation over a scaling interval $[r_1, r_2]$ with $r_2 / r_1 = 4$
         while the black curve with black error bars correspond to
         $r_2 / r_1 = 2$. The other parameters are $N=3400$, 
         $\tau_s = 0.005$ and $\rho = 10$.}
\label{fig13}
\end{figure}

Here, we use a short time series from
an epileptic seizure of $N=3400$ data sampled with $\tau_s = 0.005sec$.
The oscillations of the time series evolve irregularly, so the 
estimated $tbp \simeq 30$ does not seem to be directly related to $\tau_p$.
With a more thorough examination of the 
sequence of oscillations, we can distinguish patterns of oscillations
that may correspond to orbital periods of the potential underlying attractor. 
In Fig.\,\ref{fig13}a we show 
a part of the time series where such patterns are apparent. After severe 
filtering, the time corresponding to each pattern can be estimated by the 
{\em tbp} for the filtered time series giving $\tau_p \simeq 110$. 
Other parts of the time series
are not so regular but still patterns of about the same time length
can be identified qualitatively. 
The standard estimation procedure applied to these data gave no clear 
saturation of the $\nu$-estimate for increasing $\tau_w$, (grey curve 
in Fig.\,\ref{fig13}b). 
The estimate increases with increasing variance
showing some flatness for a small region of values of $\tau_w$ around 
$100$. In fact,
for $\tau_w > 100$ there is scaling but over a shorter interval of 
interdistances $[r1,r2]$ not satisfying the more stringent criterion 
$r2 / r1 = 4$. Relaxing this to $r2 / r1 = 2$, which has previously 
been used for EEG signals \cite{Pritchard92}, a clear saturation 
with $\nu \simeq 4$ is 
established for $\tau_w > 100$, though with increasing variance
(Fig.\,\ref{fig13}b). Thus, the optimal choice
of $\tau_w$ for the computation of $\nu$ should be around $100$, 
which is close to $\tau_p = 110$, the estimate of $\tau_w$ from  
the oscillations of the time series. Note that these results are 
not general for epileptic EEG signals.
Other EEG data showed very poor scaling and no saturation for 
increasing $\tau_w$ even for $r2 / r1 = 2$ \cite{Madsen95}
giving no valid estimate for $\nu$. In these cases, no patterns 
of oscillations could be observed. 

\section{Conclusions}

Our analysis in section 2 showed that when one reconstructs with MOD,
effective techniques for determining the delay time $\tau$ and the lowest
embedding dimension $m$ are lacking. Concerning $\tau$, there is no standard
indication of which value is the most appropriate. In fact, if we
allow $m$ to be very large, we can even use a very small $\tau$ in the
reconstruction. It seems that instead of relying on estimates 
for $\tau$ (such as the zero of the autocorrelation function or the 
minimum of mutual information) and $m$ (such as the estimate from 
false nearest neighbors)
one could rather employ ``trial and error''. In fact, this 
seems to be common in practice.

A more systematic and less tedious way to make reconstructions has been 
proposed here focusing on the time window length $\tau_w$. 
We argued that $\tau_w$ is the first 
parameter to be determined when reconstructing the state space 
and suggested that it should be approximated by the mean orbital 
period $\tau_p$. For low dimensional
attractors, $\tau_p$ is set to the time between peaks $tbp$, easily
calculated by averaging the time between successive maxima of the 
time series. Noisy time series may be filtered before determining {\em tbp}.
For higher dimensional and more complicated systems, the mean orbital period
may be found from coherent patterns of oscillations. Computationally,
this can be done measuring the ``period'' of such oscillating patterns, or 
applying strict filtering so that each pattern becomes one oscillation, and
then compute the $tbp$. 

With the estimation of $\tau_w$ and a sufficiently large $m$, the 
reconstruction is completely defined and can be used for further 
analysis of the time series. Regarding the correlation dimension, 
an initial estimate may be derived with $\tau_w=\tau_p$, and then 
checking whether the same estimate is obtained when $\tau_w$ is increased. 
For noisy data, the estimate remains the same only for $\tau_w$ close
to $\tau_p$, as noise sets an upper limit to $\tau_w$. 
The proposed parameter setting turned out to give the most confident 
$\nu$-estimates for all data analyzed where estimation was possible.

\section*{Acknowledgements}

This work has been supported by the Norwegian Research Council (NFR).
The author would like to thank Nils Christophersen for continuing 
advice and insights, and Bj{\o}rn Lillekjendlie and Torbj{\o}rn Aasen 
for their illuminative comments. The author would also like to 
thank P{\aa}l Larsson from the State Center of Epipepsy, Oslo, Norway, 
for providing the EEG data.



\end{document}